\documentclass[preprint,showpacs,superscriptaddress]{revtex4-1} 

\usepackage{amssymb}
\usepackage{subfigure}
\usepackage{amsmath}

\usepackage{graphicx}
\usepackage{epstopdf}
\usepackage{color}
\usepackage{ulem}

\newcommand{\scl}{0.2} 

\begin{document}

\title{Controllable Asymmetric Phase-Locked States of the Fundamental Active Photonic Dimer}

\date{\today}

\author{Yannis Kominis}
\affiliation{School of Applied Mathematical and Physical Science, National Technical University of Athens, Athens, Greece	}

\author{Vassilios Kovanis}
\affiliation{Department of Physics, School of Science and Technology, Nazarbayev University, Astana, Republic of Kazakhstan}

\author{Tassos Bountis}
\affiliation{Department of Mathematics, School of Science and Technology, Nazarbayev University, Astana, Republic of Kazakhstan}

\begin{abstract}
Coupled semiconductor lasers are systems possessing complex dynamics that are interesting for numerous applications in photonics. In this work, we investigate the existence and the stability of asymmetric phase-locked states  of the fundamental active photonic dimer consisting of two coupled lasers. We show that stable phase-locked states of arbitrary asymmetry exist for extended  regions of the parameter space of the system and that their field amplitude ratio and phase difference can be dynamically controlled by appropriate current injection. The model includes the important role of carrier density dynamics and shows that the phase-locked state asymmetry is related to operation conditions providing, respectively, gain and loss in the two lasers.    
\end{abstract}

\maketitle

\section{Introduction}
Coupled laser arrays are photonic structures with great potential for a large variety of applications in optical communications, sensing and imaging. One of the main features that allows such applications is their electronically controlled operation and tunability, suggesting their functionality as active metasurfaces for the transformation of appropriatelly designed spatially inhomogeneous current distributions to desirable field patterns. In that sense, a pair of coupled lasers can be considered as a fundamental element (a photonic ``molecule'') from which larger and more complicated structures can be built. The properties of such a pair of coupled lasers are determined mostly by its stationary states and their stability, that can be controlled by the current injection in the two lasers.  The existence of stable asymmetric phase-locked states with unequal field amplitudes and phase differences for the pair of coupled lasers crucially determines its far field patterns \cite{Choquette_2015, Choquette_13} and the capabilities of such a system as a building block for synthesizing larger controllable active structures characterized by complex dynamics \cite{Wang_88, Winful_90, Otsuka_90, Winful_92, Rogister_07, Soriano_13, Hizanidis_17, Erneux_book}. In addition to beam shaping applications \cite{Valagiannopoulos}, a pair of coupled lasers can be considered as an element of a ``photonic processor'' \cite{Yamamoto_1, Yamamoto_2}.\

A pair of coupled lasers is also a fundamental element for non-Hermitian optics that have been recently the subject of intense research interest. In this context, laser dynamics is commonly described by coupled mode equations for the complex field amplitudes  \cite{Choquette_2017, El-Ganainy_2016, Christodoulides_2014, Christodoulides_2016, Rotter_2012} and cases of $PT$-symmetric configurations have been considered \cite{PT_1, PT_2, PT_3, PT_4, PT_5, PT_6}. The essential condition for $PT$-symmetry in a linear system is that there is no detuning between the two lasers; in most cases balanced gain and loss are considered, however, this condition can be relaxed to an arbitrary gain or loss contrast \cite{Christodoulides_2016, Choquette_2017}. Similar studies have been reported for coupled microcavities \cite{Chong_2016} and coupled waveguides \cite{Ramezani_2010}. Deviation from exact $PT$-symmetry can be either necessitated by practical reasons \cite{Christodoulides_2016} or intentionally designed due to advantages related to the existence of stable Nonlinear Supermodes \cite{Kominis_2016, Kominis_2017}. \

The coupled mode equations, considered in the study of $PT$-symmetric lasers commonly ignore the nonlinearity of the system due to the dynamic coupling between field amplitudes and carrier densities. This approximation is valid only when the field amplitudes have small values at or below the threshold or when we take a constant gain coefficient at its saturated values, given the knowledge of the field amplitudes \cite{Choquette_2017}. More importantly, this approximation excludes some important features of the complex dynamics of the system that can be interesting with respect to photonics applications, such as the existence of symmetric and asymmetric phase-locked states and limit cycles \cite{Winful&Wang_88, Yanchuk_04} as well as localized synchronization effects \cite{Kuske&Erneux_97, Kovanis_97}, that can be described only when carrier density dynamics is taken into account \cite{Winful&Wang_88, Choquette_2017}. The latter allows for non-fixed but dynamically evolving gain and loss that enter the coupled field equations and introduce multiscale characteristics of the system due to the significant difference between carrier and photon lifetimes, resulting in dynamical features that have no counterpart in standard coupled oscillator systems \cite{Aranson_1990}. Moreover, carrier density dynamics introduces the role of current injection as a control mechanism for determining the dynamics of the system and its stationary states.   

In this work, we investigate coupled laser dynamics in terms of a model taking into account both laser coupling and carrier density dynamics. More specifically, we study the existence and stability of asymmetric phase-locked modes and investigate the role of detuning and inhomogeneous pumping between the lasers. \textit{For the case of zero detuning, we show the existence of stable asymmetric modes even when the two lasers are homogeneously pumped, i.e. the lasers are absolutely symmetric.} These modes  bifurcate to stable limit cycles in an ``oscillation death'' scenario (Bar-Eli effect) although an analogous requirement for dissimilar oscillators is not fullfiled \cite{Aranson_1990}. \textit{For the case of nonzero detuning, we show the existence of phase-locked modes with arbitrary power, amplitude ratio and phase difference for appropriate selection of pumping and detuning values. } These asymmetric states are shown to be stable for large regions of the parameter space, in contrast to common coupled oscillators where asymmetric states are usually unstable \cite{Aranson_1990}. Clearly, the above differences between coupled laser dynamics and common coupled oscillator systems are a consequence of the inclusion of carrier density dynamics in the model. In all cases, the asymmetric states have carrier densities corresponding to values that are above and below threshold resulting in gain coefficients of  opossite signs in each laser, so that the respective electric fields experience gain and loss, as in the case of $PT$-symmetric configurations. However, it is shown that deviation from $PT$-symmetry, expressed by a non-zero detuning, enables the existence of asymmetric phase-locked states of arbitrary field amplitude ratio and phase difference, that are most promising for applications.

 This paper is organized as follows: The rate equations model for coupled diode lasers is described in Section II. In Sections III and IV we analytically solve the inverse problem of determining the phase-locked states of the system, that is, we obtain the appropriate selection of the parameters of the system in order to have a given asymmetric phase-locked state and investigate their stability, for the case of zero and non-zero detuning, respectivelly. In Section V we numerically investigate the deformation of the symmetric phase-locked states of the system under nonzero detuning and asymmetric pumping. In Section VI the main conclusions of this work are summarized.

\section{Rate equations model for coupled diode lasers}
The dynamics of an array of $M$ evanescently coupled semiconductor lasers is governed by the following equations for the slowly varying complex amplitude of the normalized electric field $\mathcal{E}_i$ and the normalized excess carrier density $N_i$ of each laser:
\begin{eqnarray}
 \frac{d\mathcal{E}_i}{dt}&=&(1-i\alpha)\mathcal{E}_i N_i+i\eta(\mathcal{E}_{i+1}+\mathcal{E}_{i-1}) +i\omega_i \mathcal{E}_i \nonumber \\
 T\frac{dN_i}{dt}&=&P_i-N_i-(1+2N_i)|\mathcal{E}_i|^2 \hspace{7em} i=1...M \label{array}
\end{eqnarray}
where $\alpha$ is the linewidth enhancement factor, $\eta$ is the normalized coupling constant, $P_i$ is the normalized excess electrical pumping rate, $\omega_i$ is the normalized optical frequency detuning from a common reference, $T$ is the ratio of carrier to photon lifetimes, and $t$ is the normalized time. \cite{Winful&Wang_88} 
When the  lasers are uncoupled ($\eta=0$), they exhibit free running relaxation with frequencies
\begin{equation}
 \Omega_i=\sqrt{\frac{2P_i}{T}} 
 \end{equation}
Since we consider inhomogeneously pumped ($P_i \neq P_j$) lasers, we use a reference value $P$ in order to define a frequency 
\begin{equation}
  \Omega=\sqrt{\frac{2P}{T}} \label{Omega_0}
\end{equation}
which is further untilized in order to rescale Eqs. (\ref{array}) as
\begin{eqnarray}
 \frac{d\mathcal{E}_i}{d\tau}&=&(1-i\alpha)\mathcal{E}_i Z_i+i\Lambda(\mathcal{E}_{i+1}+\mathcal{E}_{i-1}) +i\Omega_i \mathcal{E}_i \nonumber \\
 2P\frac{dZ_i}{d\tau}&=&P_i-\Omega Z_i-(1+2\Omega Z_i)|\mathcal{E}_i|^2 \hspace{7em} i=1...M \label{array_norm}
\end{eqnarray}
where 
\begin{equation}
 \tau\equiv\Omega t, Z_i \equiv N_i/\Omega, \Lambda\equiv\eta/\Omega, \Omega_i\equiv\omega_i/\Omega
\end{equation}

In the following, we investigate the existence and stability of asymmetric phase-locked states for a pair of coupled lasers under symmetric or asymmetric electrical pumping. By introducing the amplitude and phase of the complex electric field amplitude in each laser as $\mathcal{E}_i=X_ie^{i\theta_i}$, the Eq. (\ref{array_norm}) for $M=2$ are written as
\begin{eqnarray}
 \frac{dX_1}{d\tau}&=&X_1Z_1-\Lambda X_2\sin\theta \nonumber \\
 \frac{dX_2}{d\tau}&=&X_2Z_2+\Lambda X_1\sin\theta \nonumber \\
 \frac{d\theta}{d\tau}&=&\Delta -\alpha(Z_2-Z_1)+\Lambda\left(\frac{X_1}{X_2}-\frac{X_2}{X_1}\right)\cos\theta   \label{pair} \\
2P\frac{dZ_1}{d\tau}&=&P_1-\Omega Z_1-(1+2\Omega Z_1)X_1^2 \nonumber \\
2P\frac{dZ_2}{d\tau}&=&P_2-\Omega Z_2-(1+2\Omega Z_2)X_2^2 \nonumber
\end{eqnarray}
where $\Delta=\Omega_2-\Omega_1$ is the detuning, $\theta=\theta_2-\theta_1$ is the phase difference of the electric fields and we have used a reference value $P=(P_1+P_2)/2$ in order to define $\Omega$ as in Eq. (\ref{Omega_0}). As a reference case, we consider a pair of lasers with $\alpha=5$, $T=400$ which is a typical configuration relevant to experiments and we take $P=0.5$. For these values we have $\Omega=5 \times 10^{-2}$ and a coupling constant $\eta$ in the range of $10^{-5} \div 10^0$ corresponds to a $\Lambda$ in the range $0.5 \times 10^{-3} \div 0.5 \times 10^{2}$.
The phase-locked states are the equilibria of the dynamical system (\ref{pair}), given as the solutions of the algebraic system obtained  by setting the time derivatives of the system equal to zero and their linear stability is determined by the eigenvalues of the Jacobian of the system. For the case of zero detuning ($\Delta =0$) and symmetric electrical pumping ($P_1=P_2=P_0$), two phase-locked states are known analytically: $X_1=X_2=\sqrt{P_0}$, $Z_1=Z_2=0$ and $\theta=0,\pi$. The in-phase state ($\theta=0$) is stable for $\eta>\alpha P_0 /(1+2P_0)$ whereas the out-of-phase state ($\theta=\pi$) is stable for $\eta<(1+2P_0)/2\alpha T$ \cite{Winful&Wang_88}. 

The phase difference $\theta$ and the electric field amplitude ratio $\rho \equiv X_2/X_1$ of a phase-locked state crucially determine the intensity response of the system. The incoherent intensity is defined as the sum of the individual laser intensities $S \equiv |\mathcal{E}_1|^2+|\mathcal{E}_2|^2=(1+\rho^2)X_1^2$ and can be measured by placing a broad detector next to the output face of the system. The coherent intensity corresponds to a coherent superposition of the individual electric fields $I \equiv |\mathcal{E}_1+\mathcal{E}_2|^2=(1+\rho^2+2\rho \cos \theta)X_1^2$ and can be measured by placing a detector at the focal plane of an external lens. The coherent intensity depends on the phase difference $\theta$; however it does not take into account the spatial distribution of the lasers, i.e. the distance between them. The latter determines the far-field pattern of the intensity resulting from constructive and destructive interference at different directions in space. By considering, for the sake of simplicity, the lasers as two point sources at a distance $d$, the coherent intensity is given by $I_\phi=|\mathcal{E}_1+e^{i 2\pi (d/ \lambda)\sin \phi} \mathcal{E}_2 |^2=\mathcal{E}_2|=[1+\rho^2+2\rho \cos (\theta+2\pi (d/\lambda) \sin \phi)]X_1^2$ where $\phi$ is the azimuthal angle measured from the direction normal to the distance between the lasers and $\lambda$ is the wavelength. \cite{Hecht} It is clear that the asymmetry of the phase-locked state described by $\rho$ and $\theta$ along with the geometric parameter of the system $d/\lambda$ define a specific far-field pattern $I_\phi$ with desirable characteristics.

\begin{figure}[pt]
  \begin{center}
  \subfigure[]{\scalebox{\scl}{\includegraphics{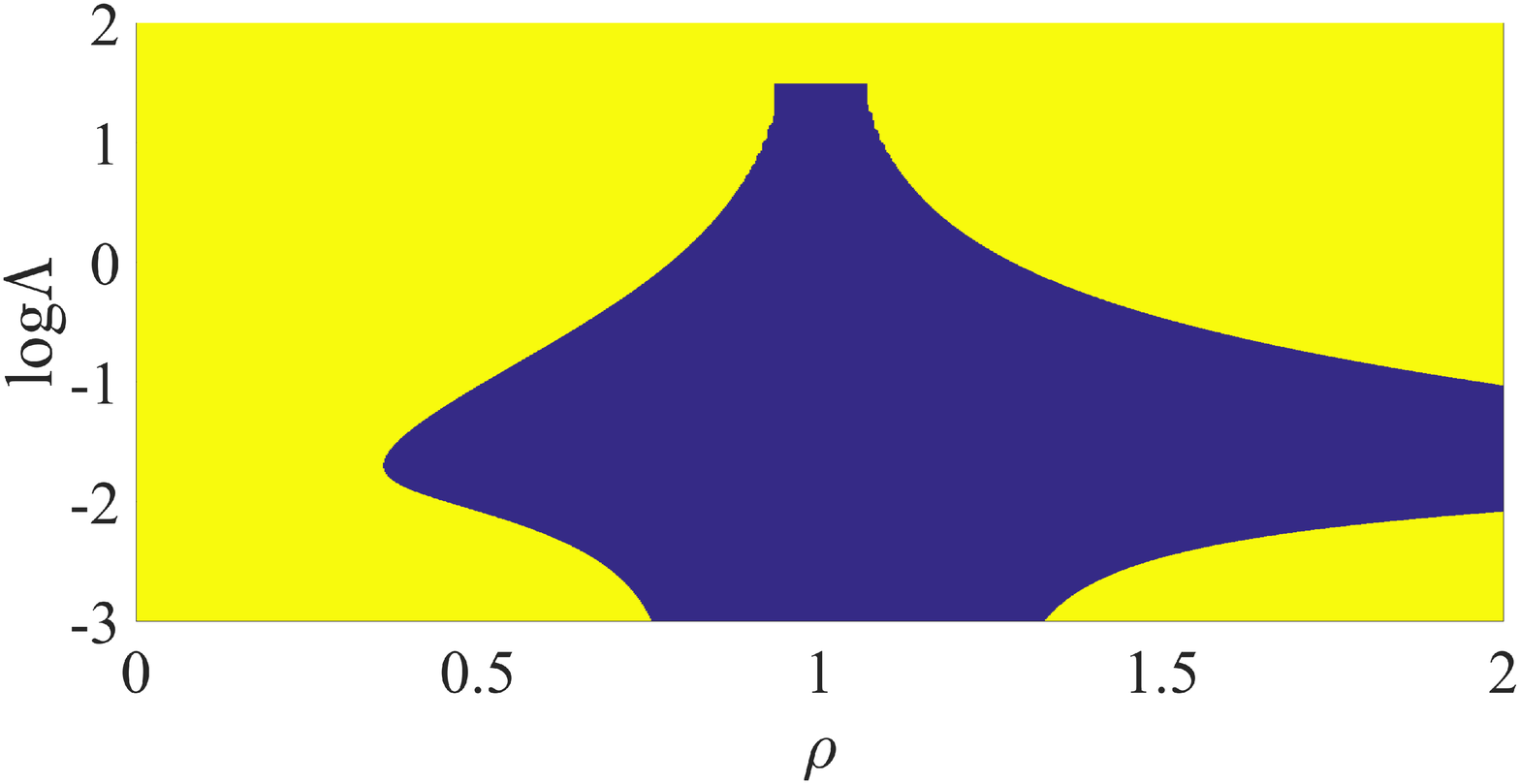}}}
  \subfigure[]{\scalebox{\scl}{\includegraphics{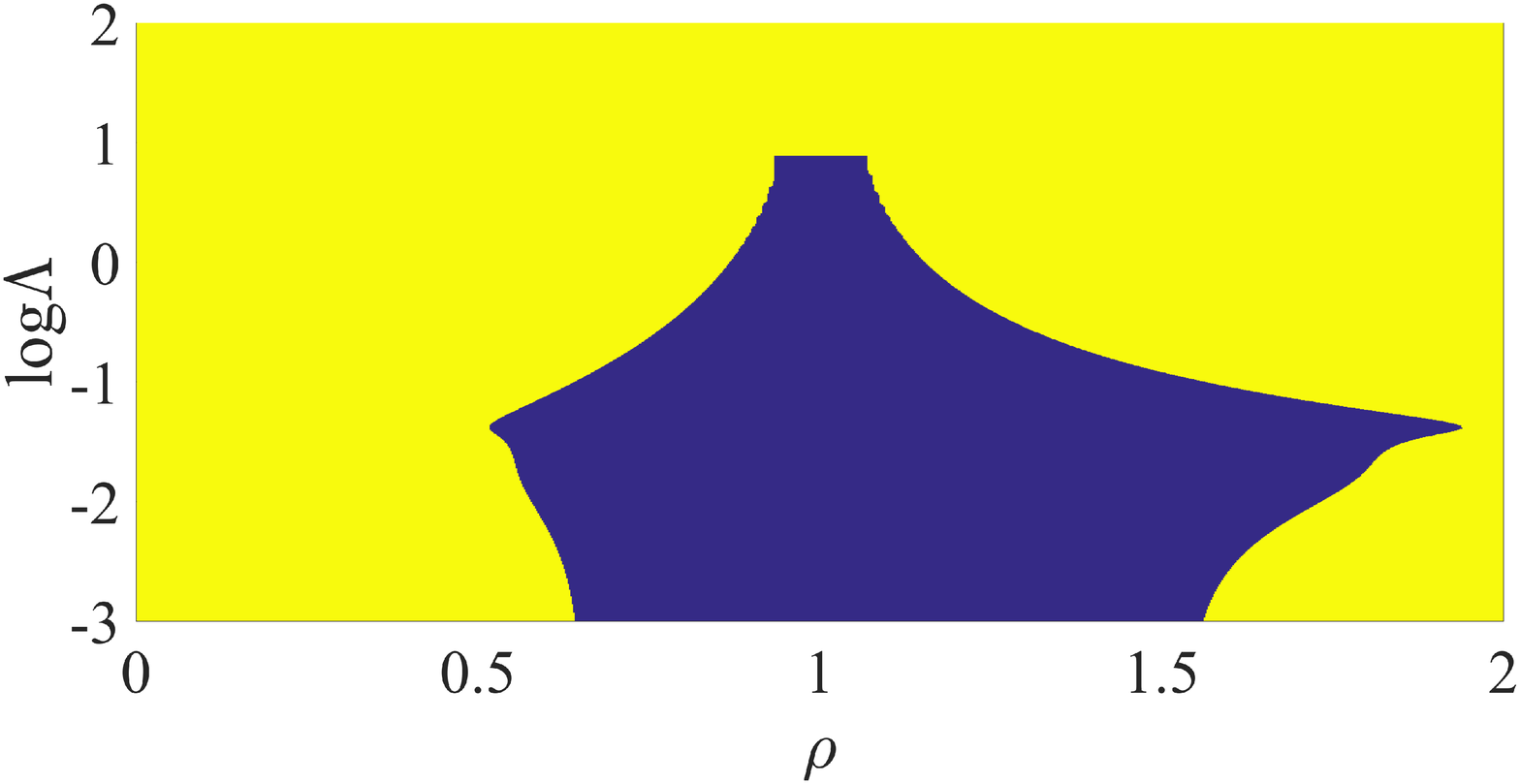}}}
  \subfigure[]{\scalebox{\scl}{\includegraphics{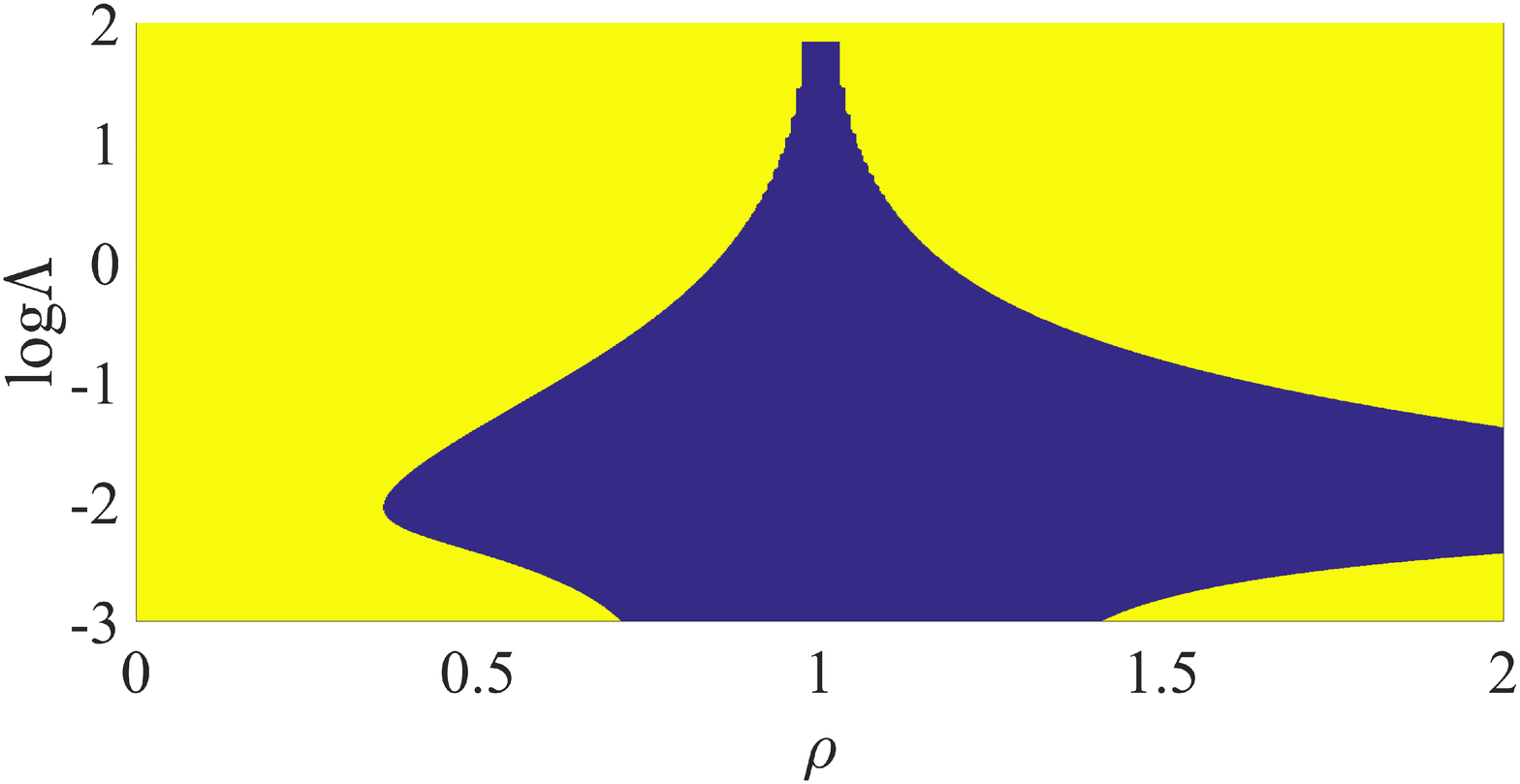}}}
  \caption{Stability regions of asymmetric phase-locked states in a symmetric configuration with $\Delta=0$ and $P_1=P_2=P_0$ in the $(\Lambda, \rho)$ parameter space. Dark blue and light yellow areas correspond to stability and instability, respectively. (a) $\alpha=5$ and $T=400$, (b)  $\alpha=1.5$ and $T=400$, (c)  $\alpha=5$ and $T=2000$ (case of \cite{Winful&Wang_88}).}
  \end{center}
\end{figure}

\begin{figure}[pt]
  \begin{center}
  \subfigure[]{\scalebox{\scl}{\includegraphics{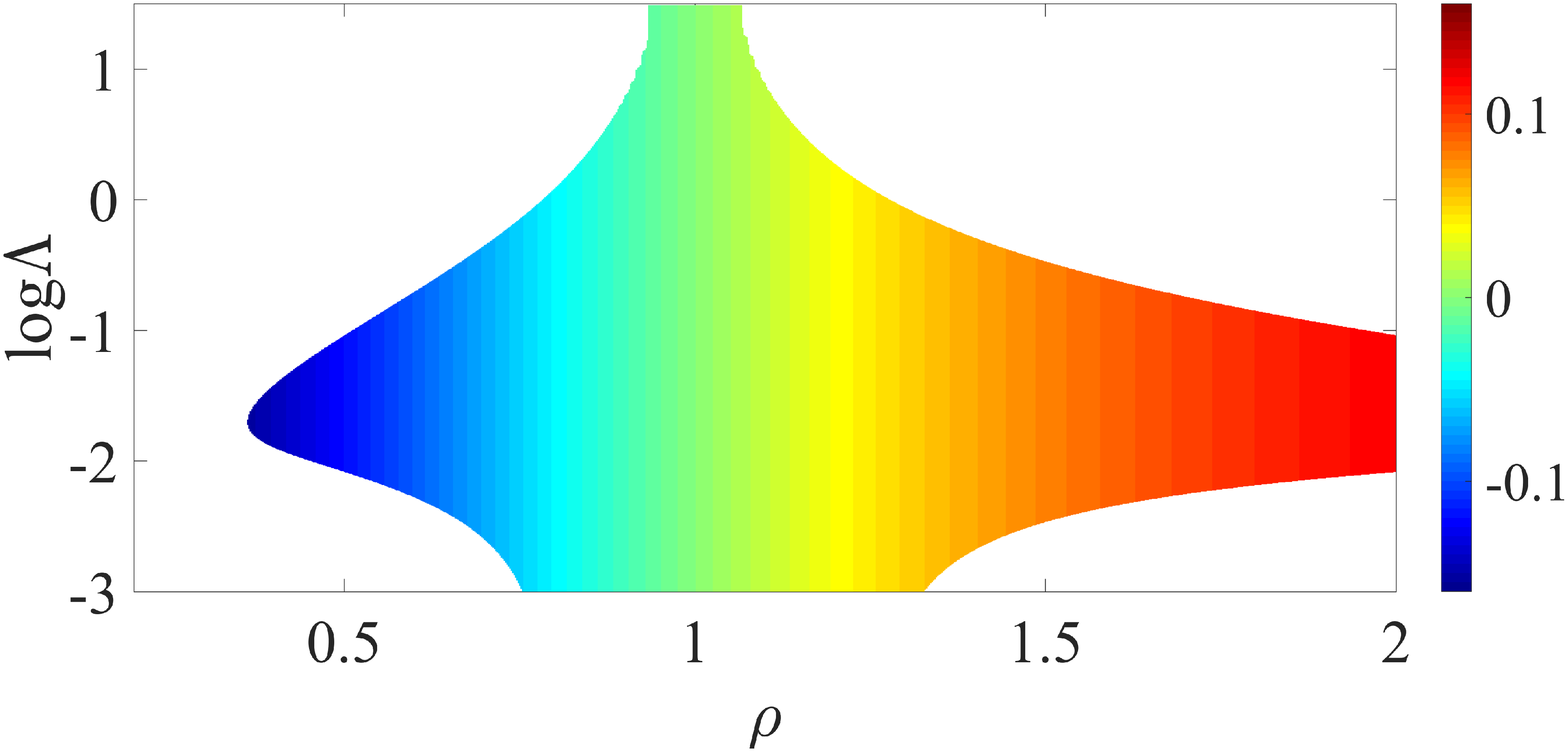}}}
  \subfigure[]{\scalebox{\scl}{\includegraphics{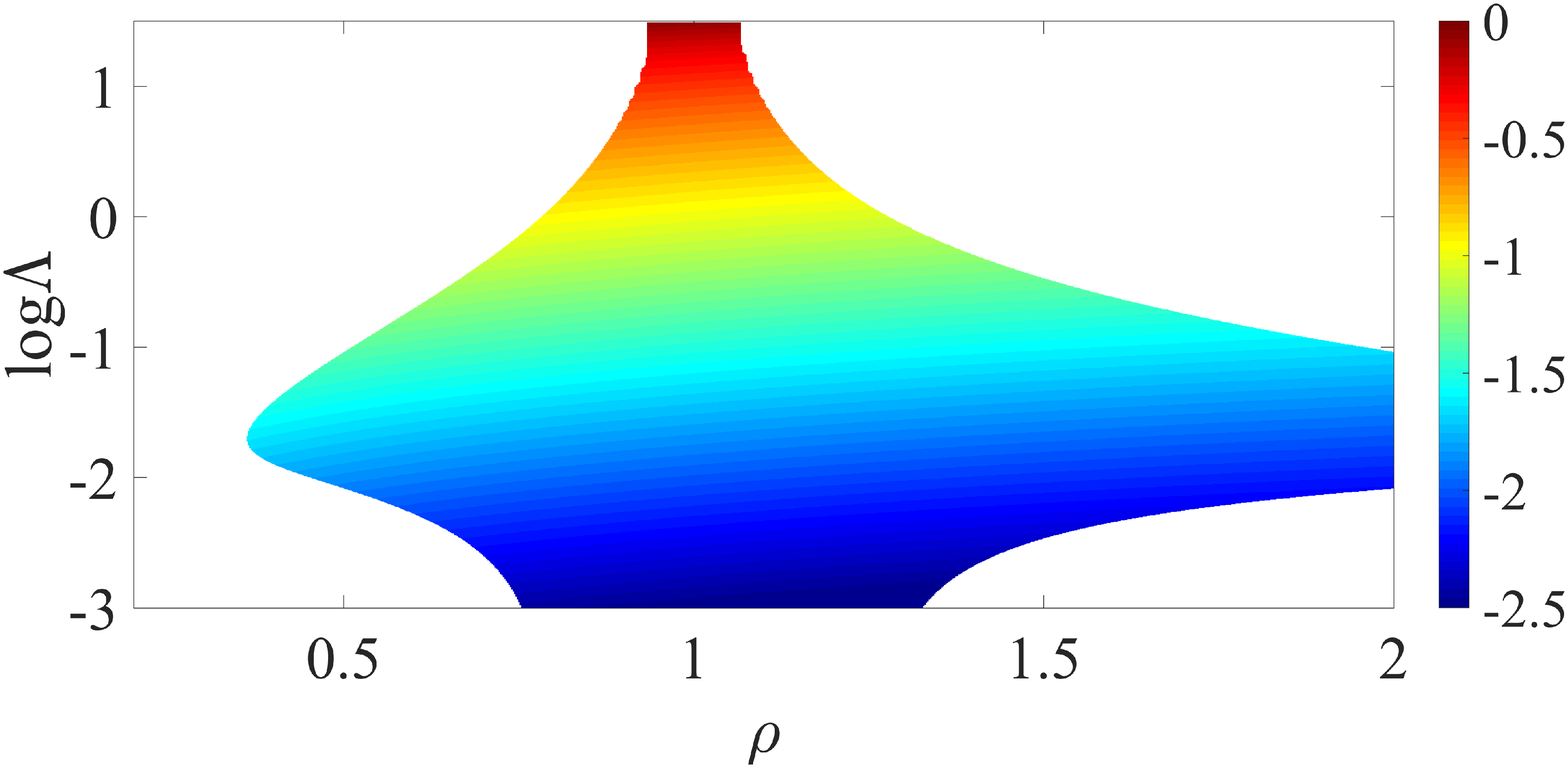}}}
  \caption{Steady-state phase difference $\theta$ (a) and field amplitude $X_0$ (in logarithmic scale) (b) of the stable asymmetric phase-locked state in the $(\Delta, \rho)$ parameter space. Parameter values correspond to Fig. 1(a).}
  \end{center}
\end{figure}

\section{Asymmetric phase-locked states under zero detuning}
Although we cannot find analytical solutions of the system of equations that provide the field amplitudes and phase difference for a given set of laser parameters, we can solve explicitly the reverse problem: for a given phase-locked state with field amplitude ratio $\rho \equiv X_2/X_1$ and phase difference $(\theta)$ we can analytically solve the algebraic system of equations obtained by setting the rhs of Eq. (\ref{pair}) equal to zero to determine the steady-state carrier densities $(Z_{1,2})$, and the appropriate detuning $(\Delta)$ and pumping rates $(P_{1,2})$, in terms of $\rho$ and $\theta$.  In this section, we consider the case of zero detuning ($\Delta=0$) between the coupled lasers. It is straightforward to verify that for every $\rho$ there exists an equilibrium of the dynamical system (\ref{pair}) with a fixed phase difference $\theta$
\begin{equation}
 \tan\theta= \frac{1}{\alpha} \frac{\rho^2-1}{\rho^2+1} \label{theta_rho} \\ 
\end{equation}
and
\begin{eqnarray}
 Z_1&=&\Lambda \rho \sin\theta \nonumber \\
 Z_2&=&-\frac{\Lambda}{\rho} \sin\theta\\
 P_1&=&X_0^2+(1+2X_0^2)\Omega\Lambda\rho\sin\theta  \nonumber \\
 P_2&=&\rho^2X_0^2-(1+2\rho^2 X_0^2) \frac{\Omega\Lambda}{\rho}\sin\theta \label{P12}  
 \end{eqnarray}
 where $X_0 \equiv X_1$.
 For the case of symmetrically pumped lasers ($P_1=P_2=P_0$) the phase-locked states have a fixed field amplitude $X_0$ given by 
\begin{equation}
X_0^2 = \frac{\Omega \Lambda \sin\theta (\rho^2+1)}{\rho\left[(\rho^2-1)-4\Omega \Lambda \rho \sin\theta\right]} \label{X0_rho}
\end{equation}
and the common pumping is 
\begin{equation}
P_0=X_0^2+(1+2X_0^2)\Omega \Lambda \rho \sin\theta \label{P0_rho}
\end{equation}

It is quite remarkable that an asymmetric phase-locked state, with arbitrary amplitude ratio ($\rho$) exists even for the case of identical coupled lasers. This asymmetric state describes a localized synchronization effect \cite{Kuske&Erneux_97, Kovanis_97}, with the degree of localization determined by the field amplitude ratio $\rho$. More interestingly, this state is stable within a large area of the parameter space as shown in Fig. 1, in contrast to what is expected \cite{Yanchuk_04}. In fact, there exist areas of the parameter space where the previously studied \cite{Winful&Wang_88} symmetric, in-phase and out-of-phase, states  are unstable, so that this asymmetric state is the only stable phase-locked state of the system. The stability of this state depends strongly on the parameters $\alpha$ and $T$, as shown in Fig. 1. The asymmetric phase-locked state undergoes Hopf-bifurcations giving rise to stable limit cycles where the fields oscillate around the respective phase-locked values, similarly to the case of symmetric phase-locked states,  \cite{Winful&Wang_88} but with different oscillation amplitudes in general \cite{Kovanis_97}.  It is worth mentioning, that this ``oscillation death'' (or Bar-Eli) effect, \cite{Aranson_1990} commonly occuring for coupled dissimilar oscillators, is taking place even for the case of identical lasers due to the consideration of the role of carrier density dynamics. 

The amplitude ratio $\rho$ determines the phase difference $\theta$ as shown in Eq. (\ref{theta_rho}) and Fig. 2(a). The extreme values of the phase difference are determined by the linewidth enhancement factor $\alpha$ with smaller values of $\alpha$ allowing for larger phase differences. Moreover, the amplitude of the field depends strongly on the amplitude ratio $\rho$ and the normalized coupling constant $\Lambda$ as shown in Eq. (\ref{X0_rho}) and Fig. 2(b). The appropriate symmetric pumping $P_0$ in order to have an asymmetric phase-locked state is given by Eq. (\ref{P0_rho}). It is obvious from the above equations that for $\rho=1$ the well-known symmetric states are obtained \cite{Winful&Wang_88}.

\begin{figure}[pt]
  \begin{center}
  \subfigure[]{\scalebox{\scl}{\includegraphics{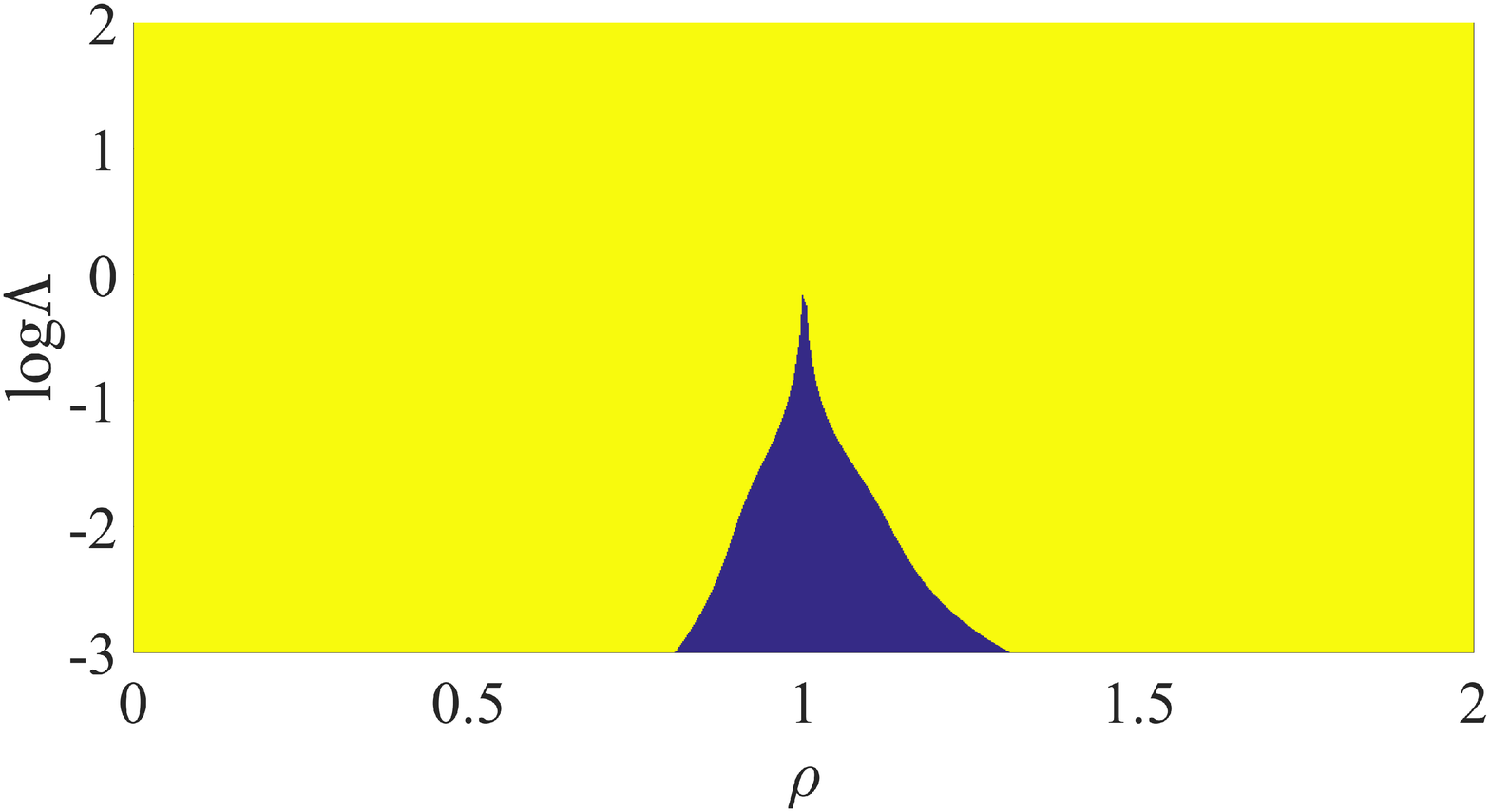}}}
  \subfigure[]{\scalebox{\scl}{\includegraphics{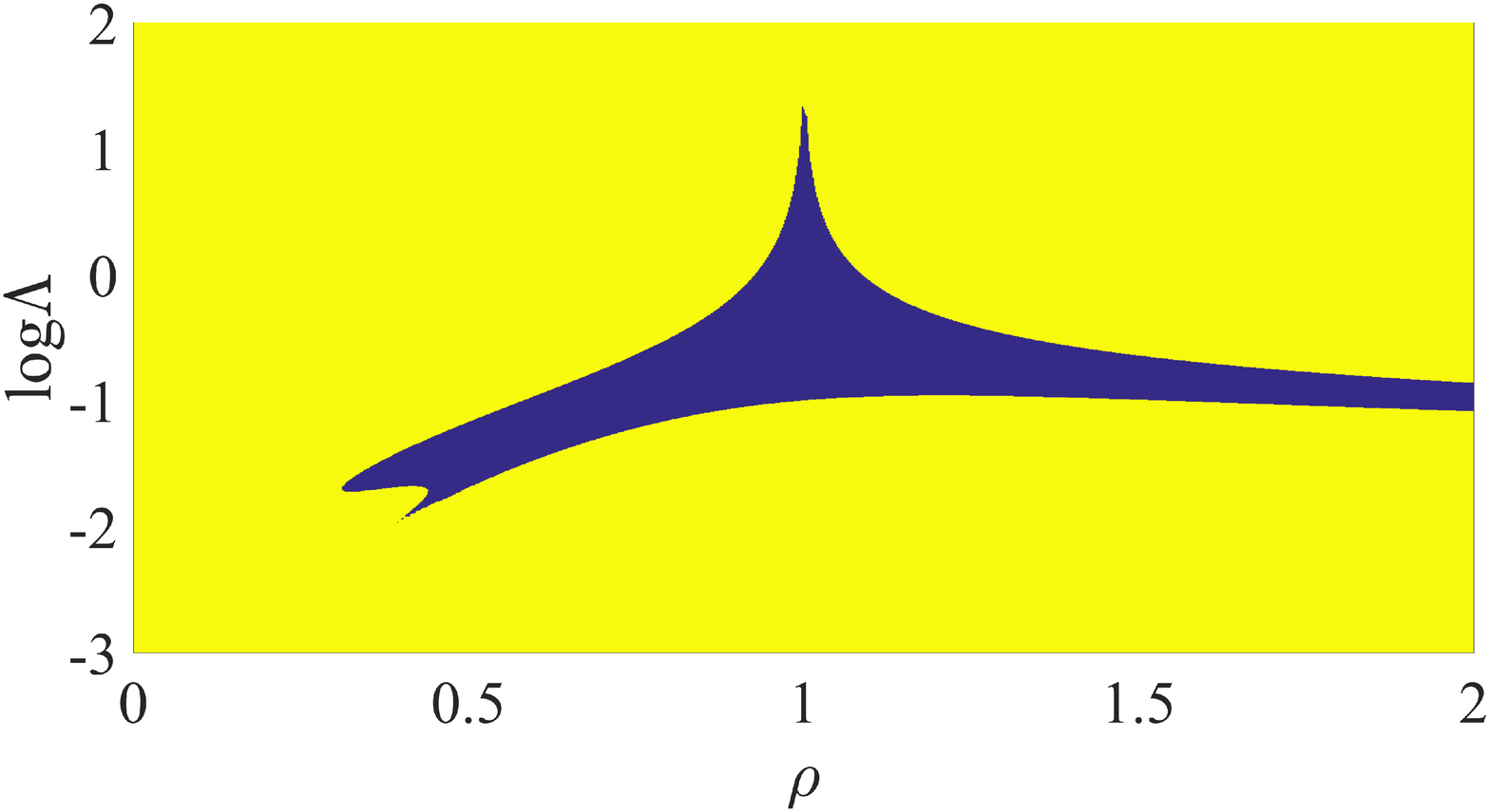}}}
  \subfigure[]{\scalebox{\scl}{\includegraphics{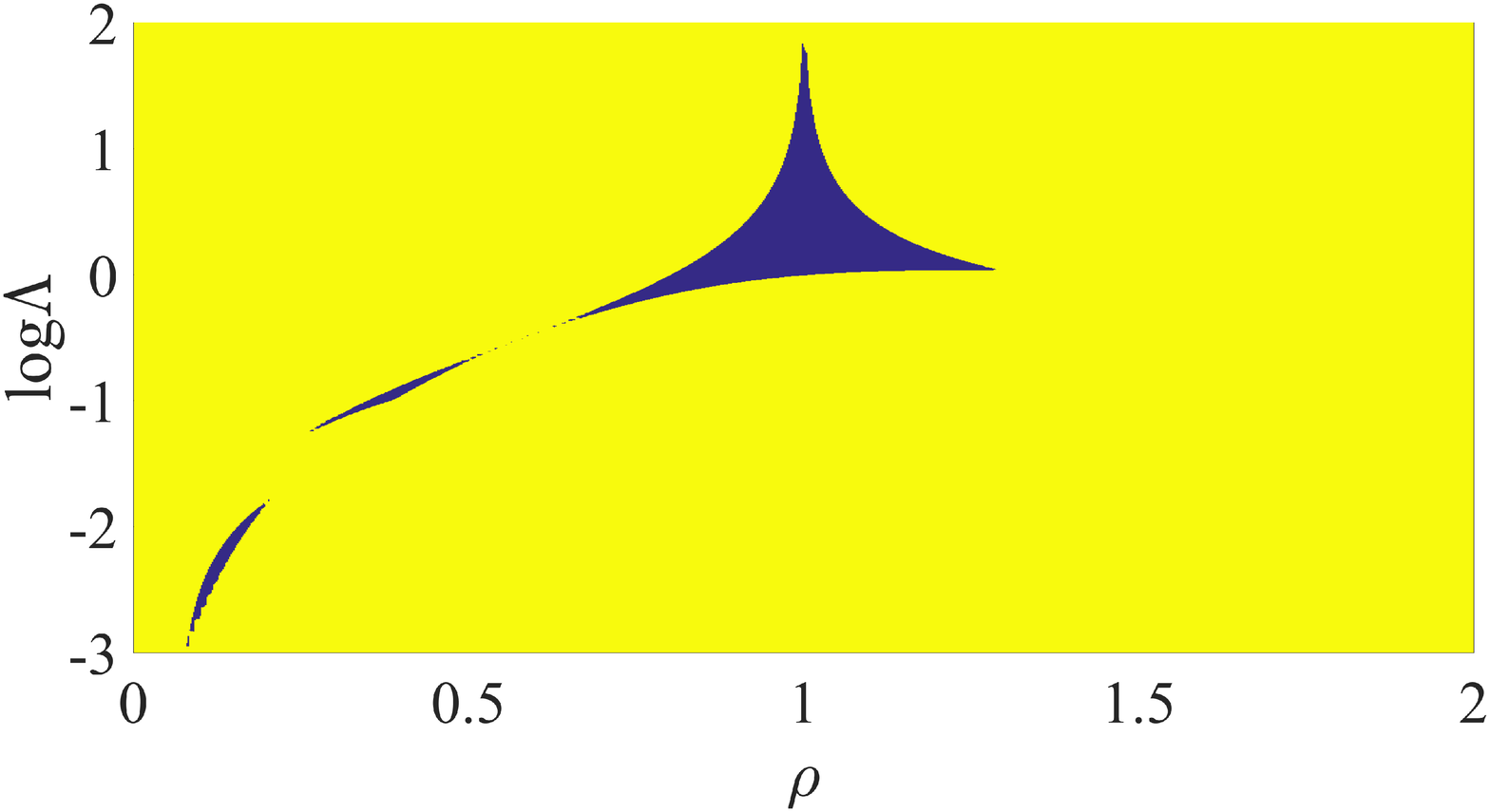}}}
  \caption{Stability regions of asymmetric phase-locked states in an asymmetric configuration with $\Delta=0$ and $P_1 \neq P_2$ in the $(\Lambda, \rho)$ parameter space. Dark blue and light yellow areas correspond to stability and instability, respectively. $\alpha=5$, $T=400$, and (a) $\log X_0=-3$, (b)  $\log X_0=-1.5$, (c) $\log X_0=-1$ .}
  \end{center}
\end{figure}

Phase-locked states with fixed phase difference $\theta$ but arbitrary electric field amplitude $X_0$ exist for different pumping between the two lasers, given by Eq. (\ref{P12}). The stability of these states depends crucially on the electric field amplitude $X_0$ as shown in Fig. 3. In comparison to Fig. 1(a) corresponding to the same parameter set but with $P_1=P_2=P_0$ the extent of the stability region is significantly reduced. 

\begin{figure}[pt]
  \begin{center}
  \subfigure[]{\scalebox{\scl}{\includegraphics{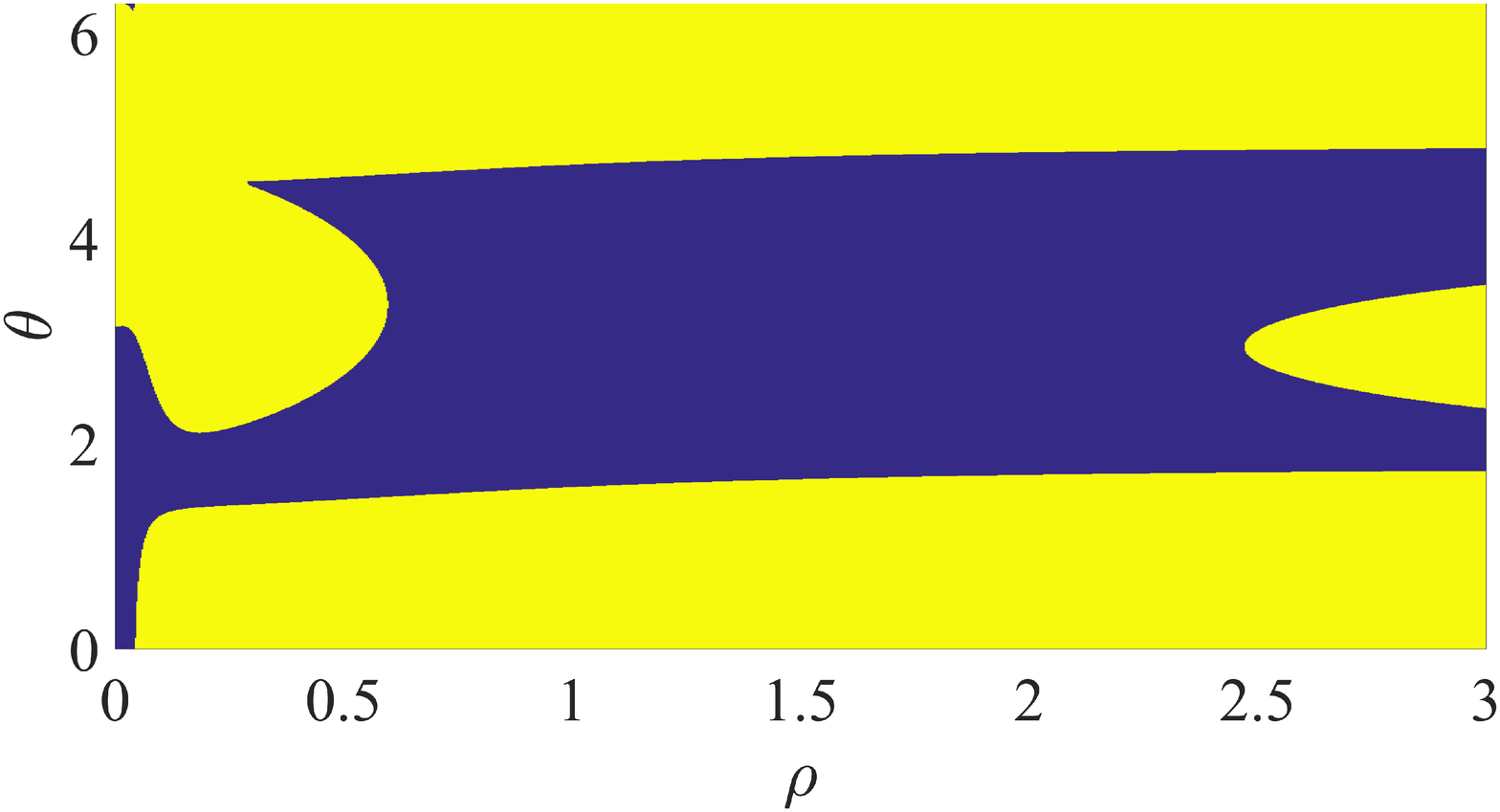}}}
  \subfigure[]{\scalebox{\scl}{\includegraphics{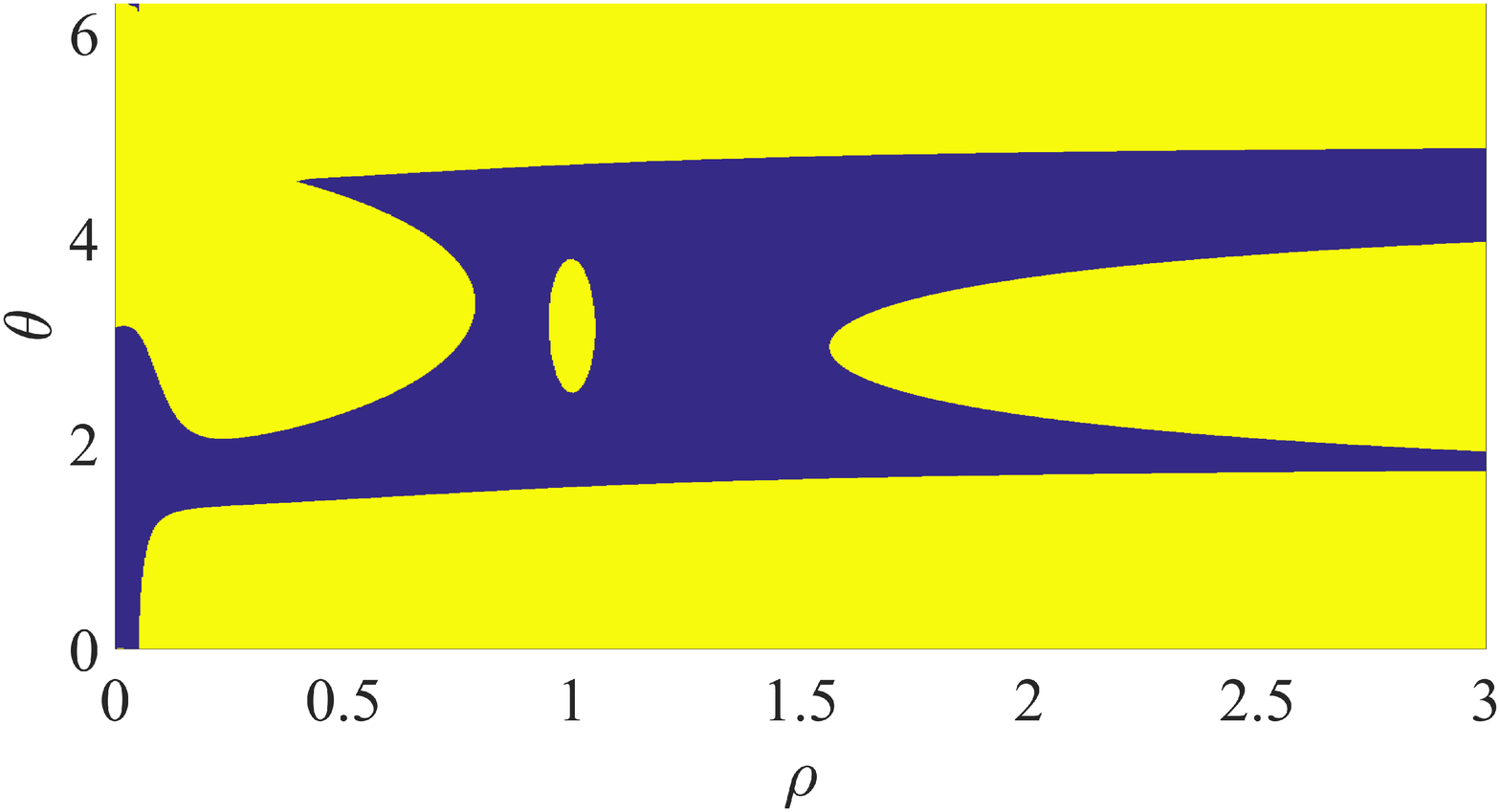}}}\\
  \subfigure[]{\scalebox{\scl}{\includegraphics{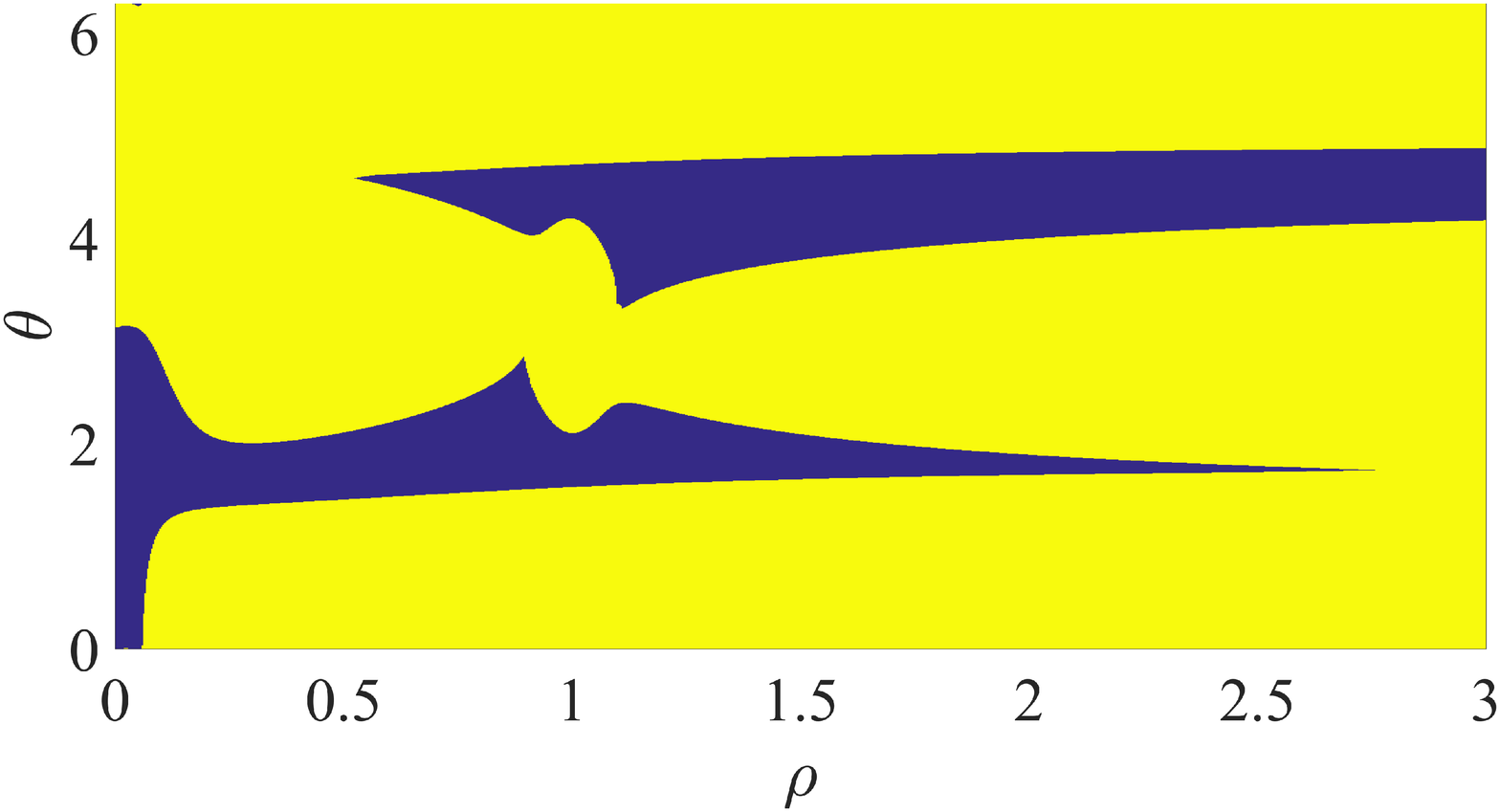}}}
  \subfigure[]{\scalebox{\scl}{\includegraphics{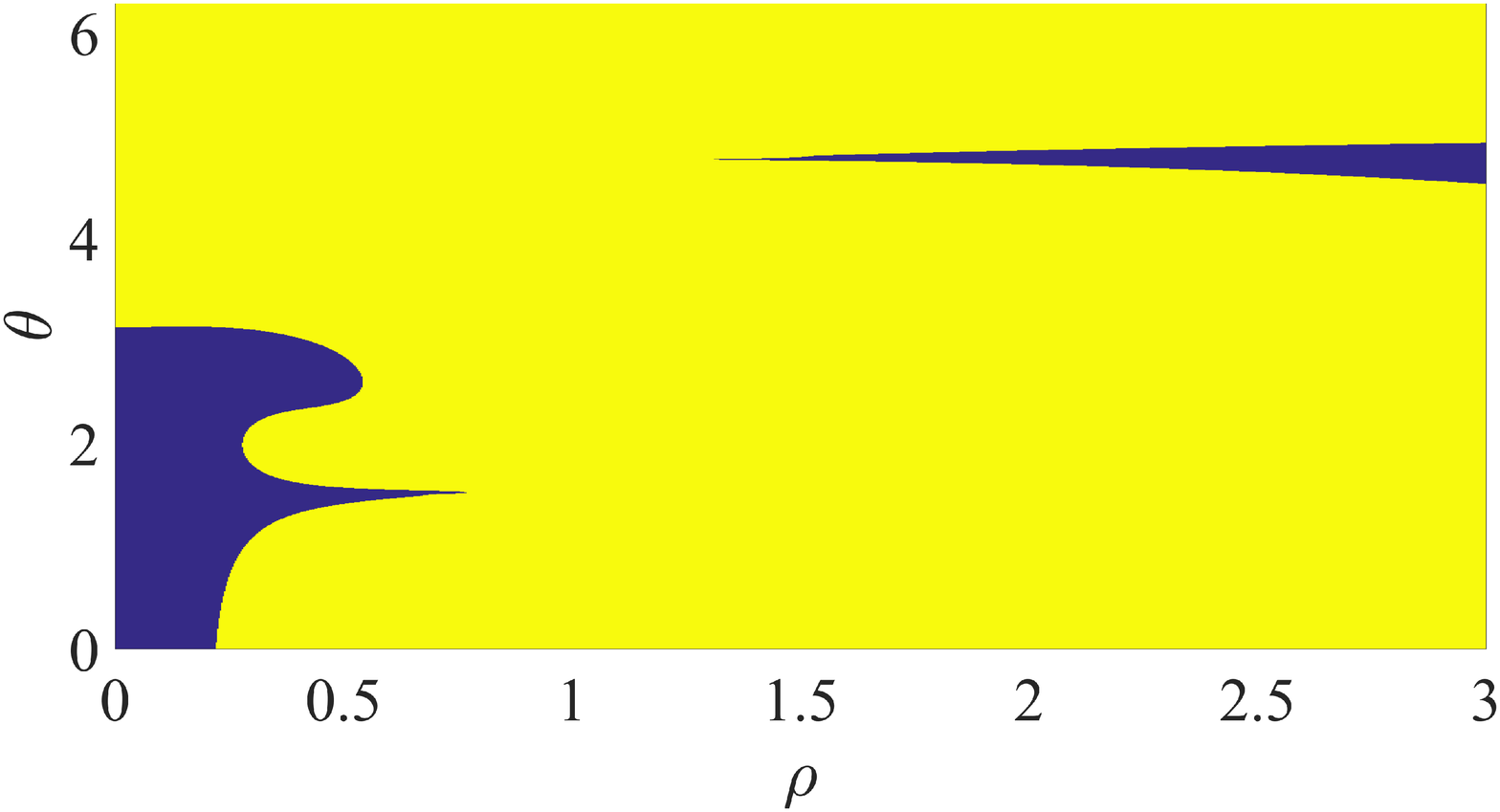}}}\\
  \subfigure[]{\scalebox{\scl}{\includegraphics{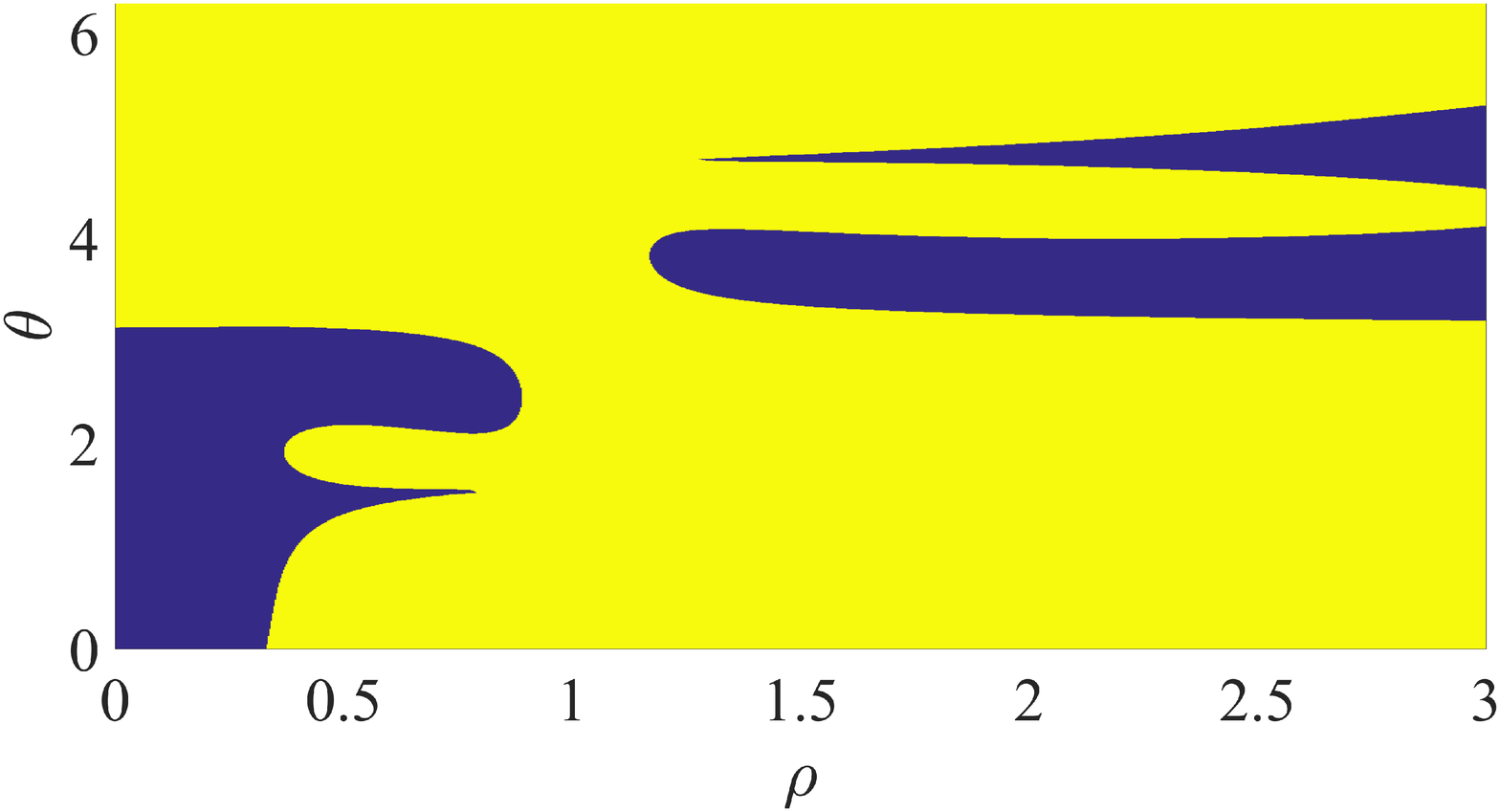}}}
  \subfigure[]{\scalebox{\scl}{\includegraphics{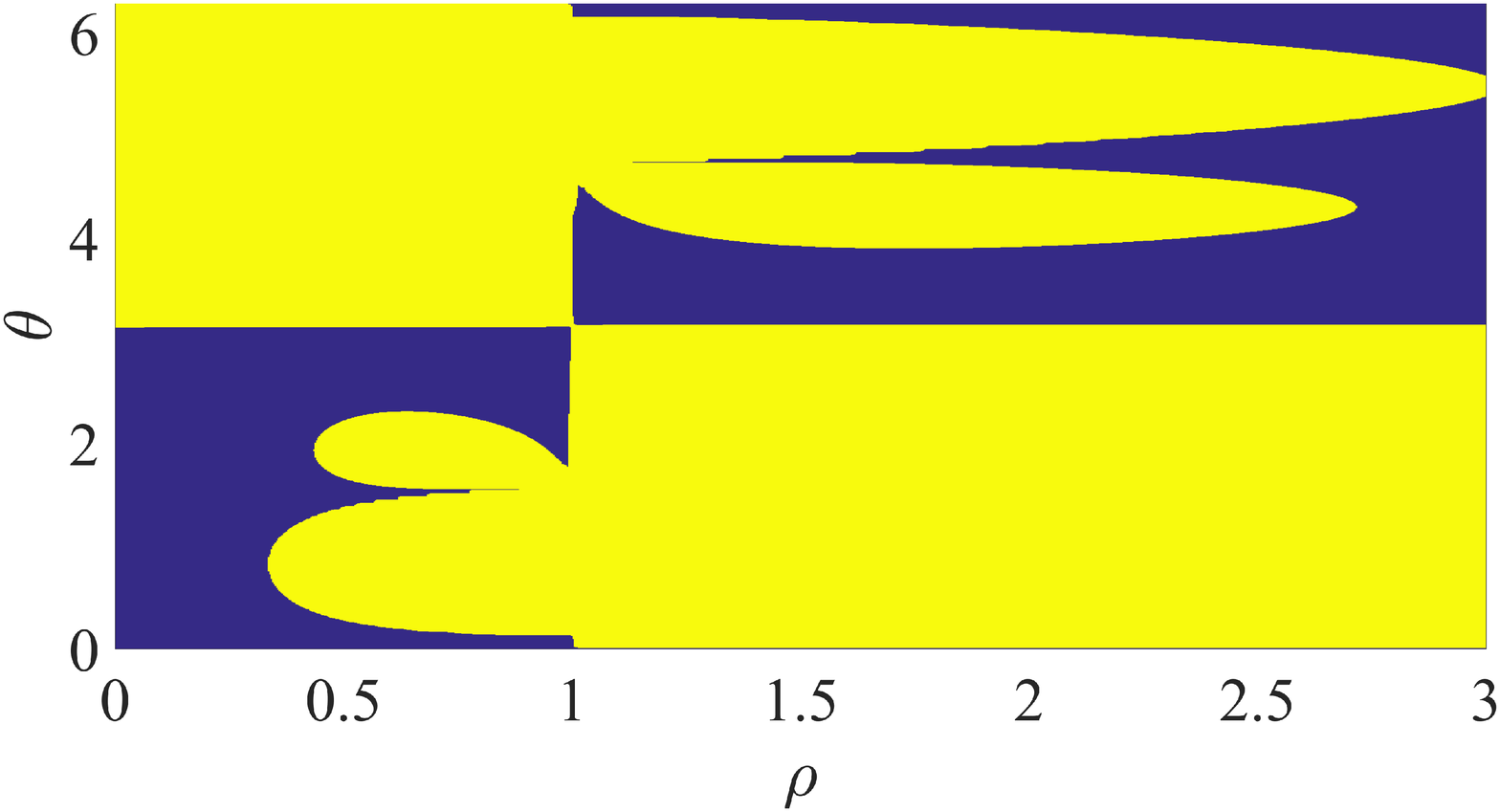}}}
   \caption{Stability regions of phase-locked states of arbitrary asymmetry, characterized by the steady-state field amplitude ratio $\rho$, phase difference $\theta$ and $X_0=\sqrt{0.5}$. Dark blue and light yellow areas correspond to stability and instability, respectively. The parameters of the coupled lasers are $\alpha=5, T=400$ and $\log\Lambda=-2.1,-1.9,-1.7,0,0.5,2$ (a)-(f). The respective detuning $\Delta$ and pumping $P_{1,2}$ values are given by Eqs. (\ref{D_eq})-(\ref{P_eq}) . The stability regions are symmetric with respect to the transformation $\rho \rightarrow 1/\rho$ and $\theta \rightarrow 2\pi-\theta$. The topology and the extent of the stability region depends crucially on the coupling coefficient $\Lambda$. }
  \end{center}
\end{figure}

\begin{figure}[pt]
  \begin{center}
  \subfigure[]{\scalebox{\scl}{\includegraphics{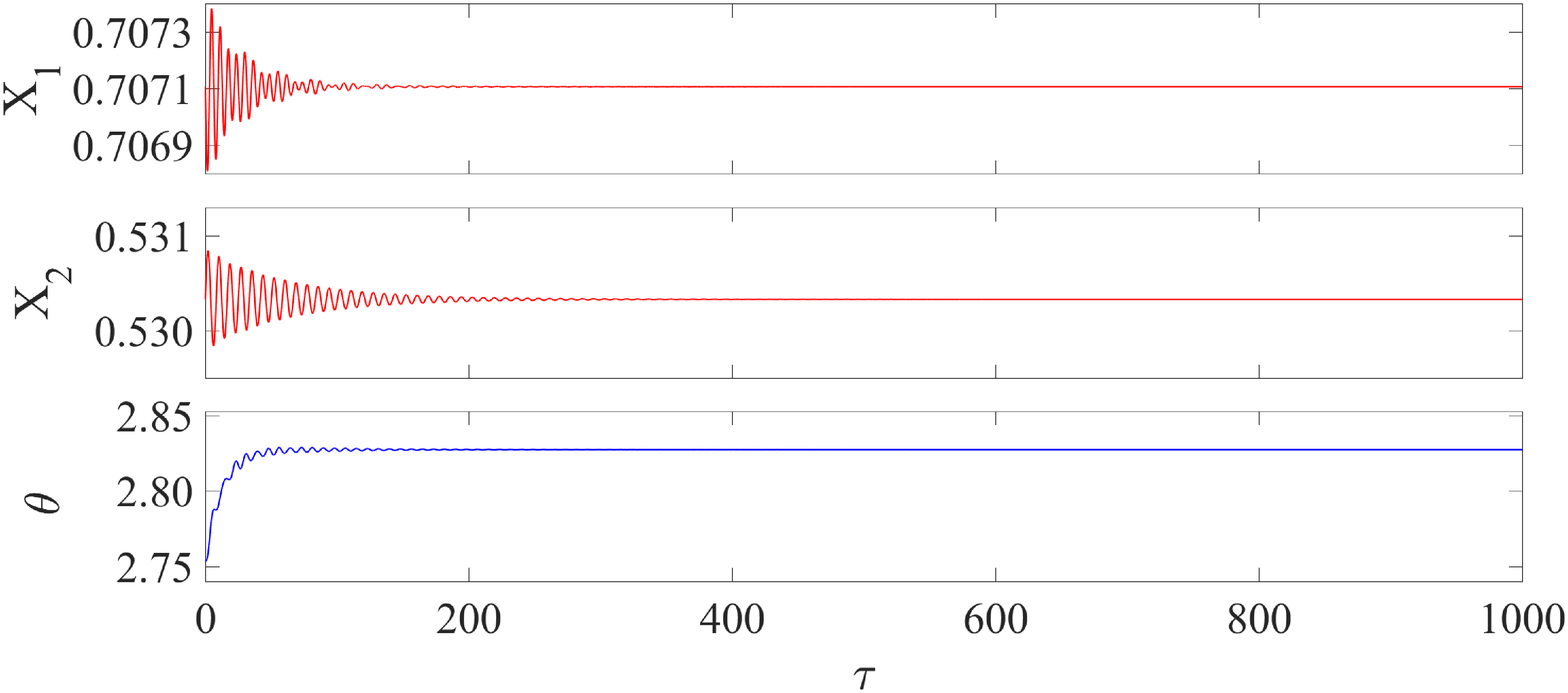}}}
  \subfigure[]{\scalebox{\scl}{\includegraphics{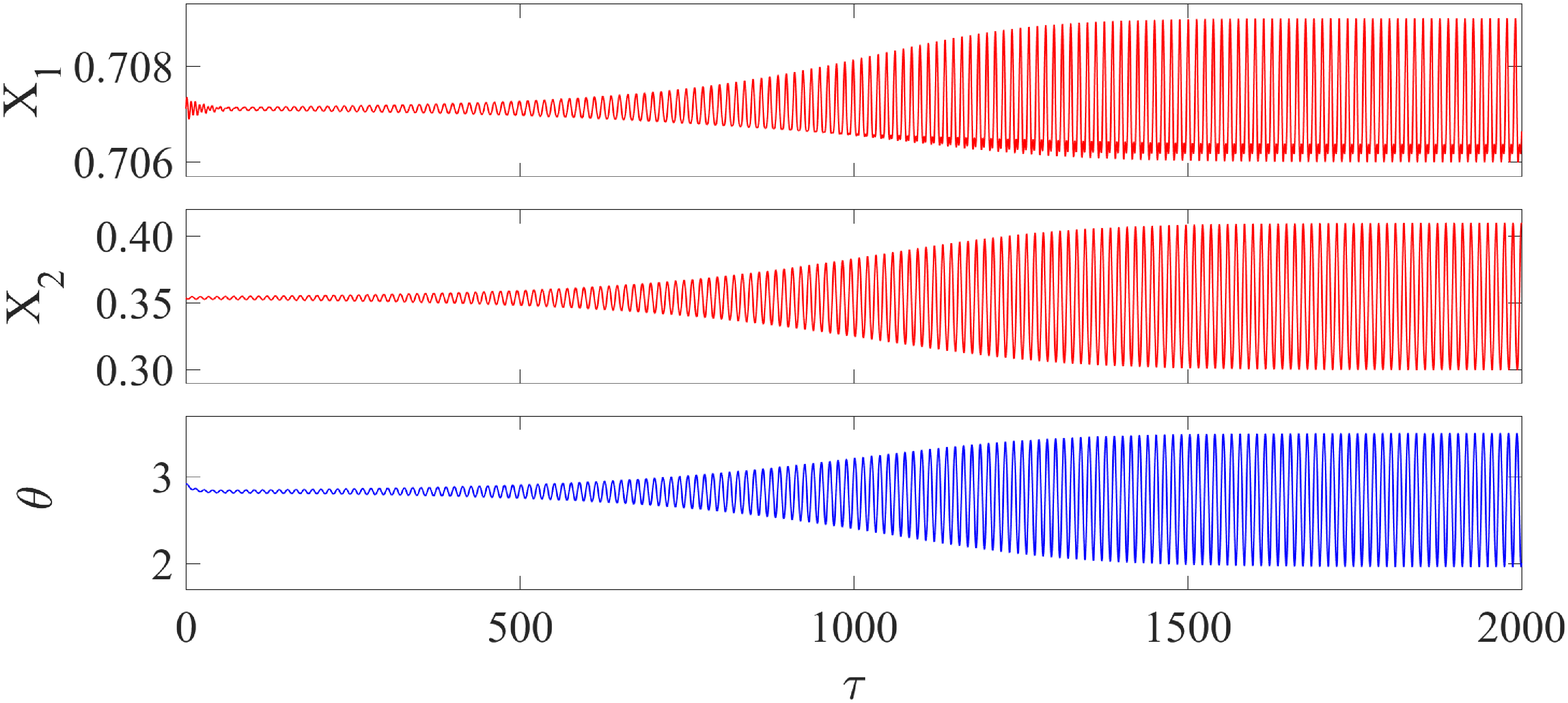}}}\\
  \subfigure[]{\scalebox{\scl}{\includegraphics{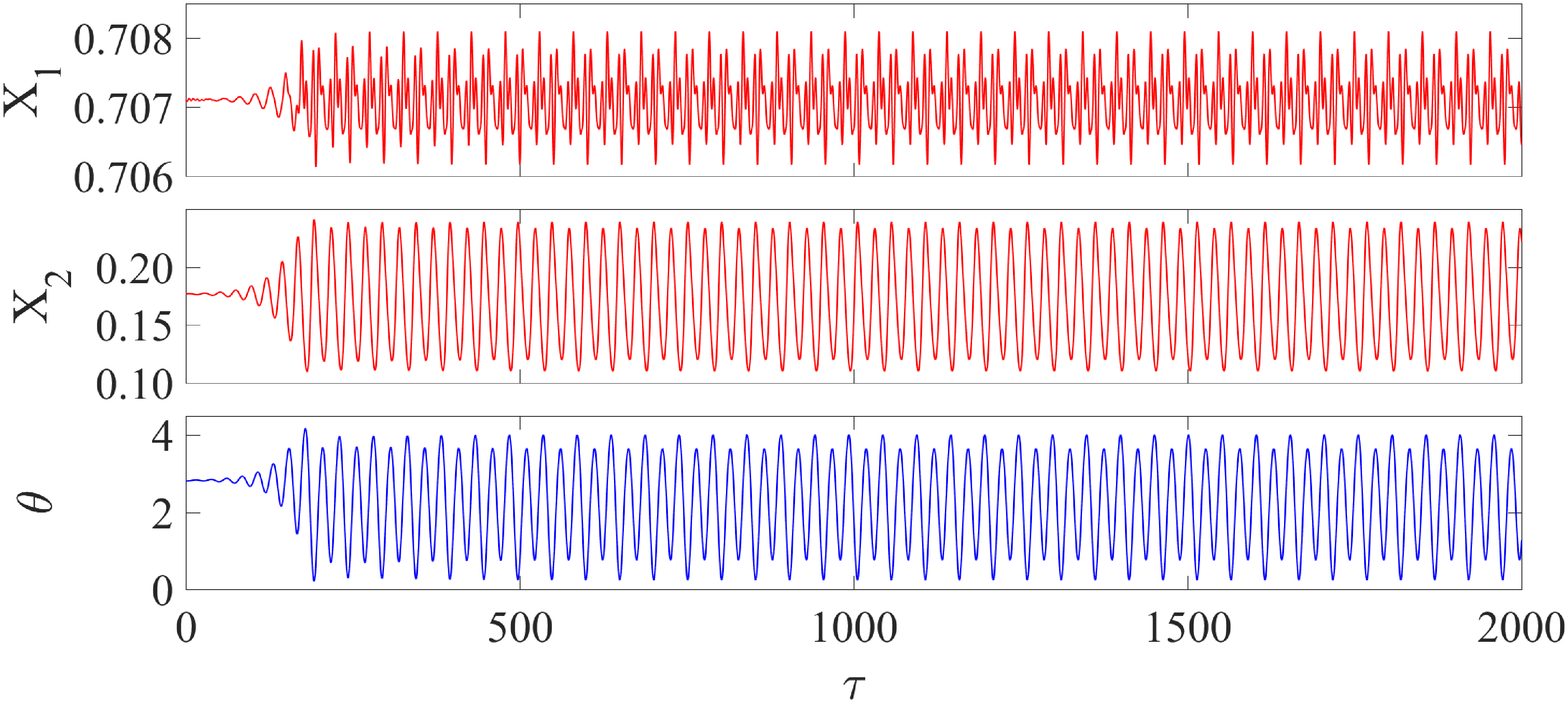}}}
  \subfigure[]{\scalebox{\scl}{\includegraphics{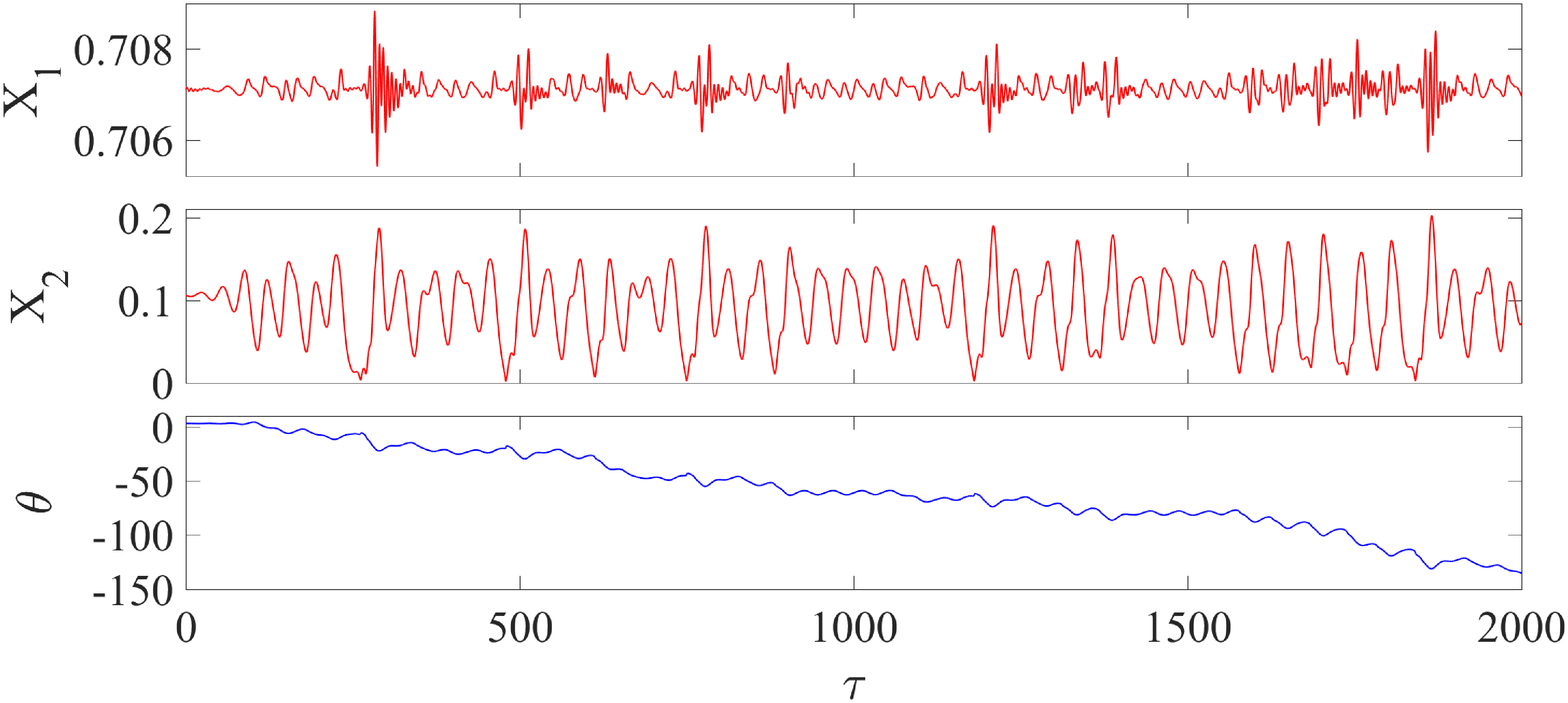}}}\\
  \subfigure[]{\scalebox{\scl}{\includegraphics{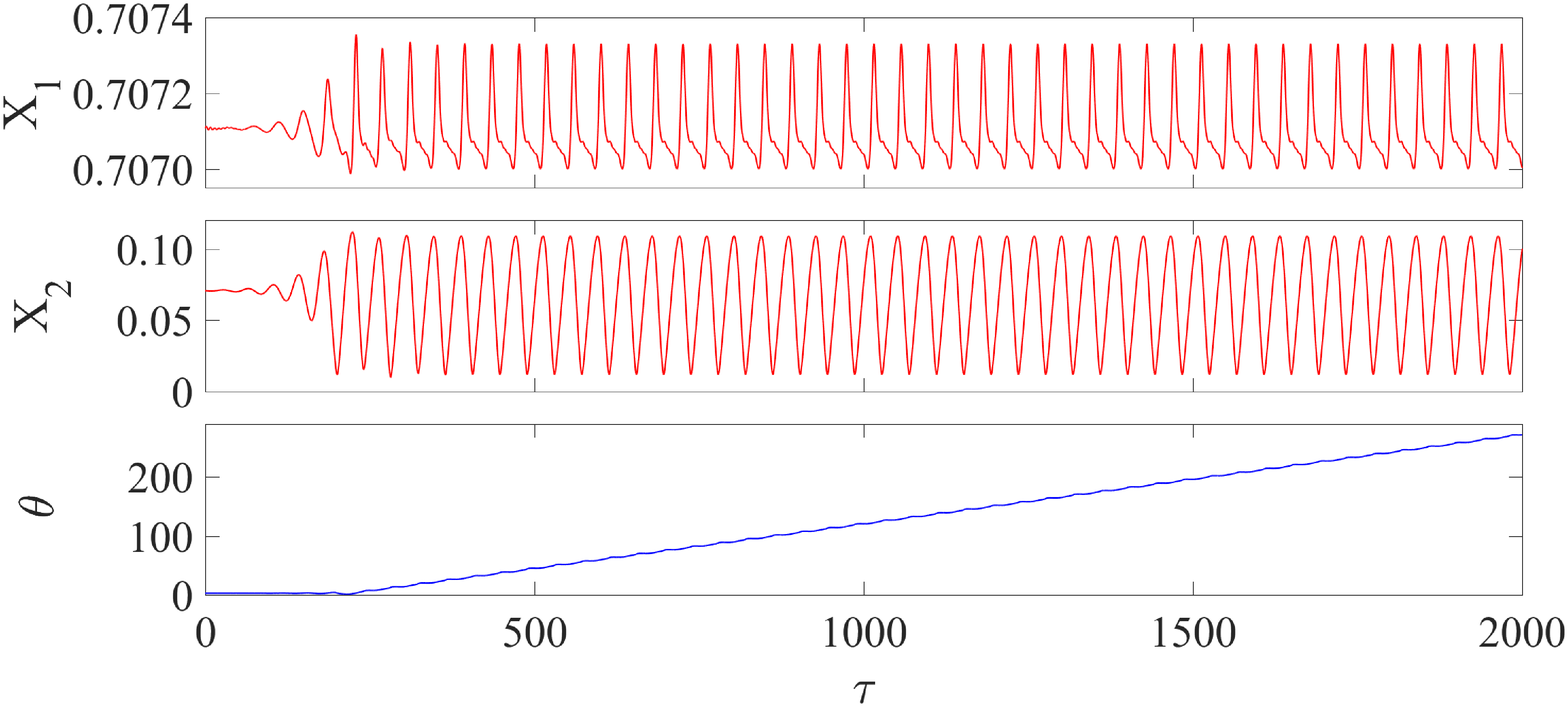}}}
  \subfigure[]{\scalebox{\scl}{\includegraphics{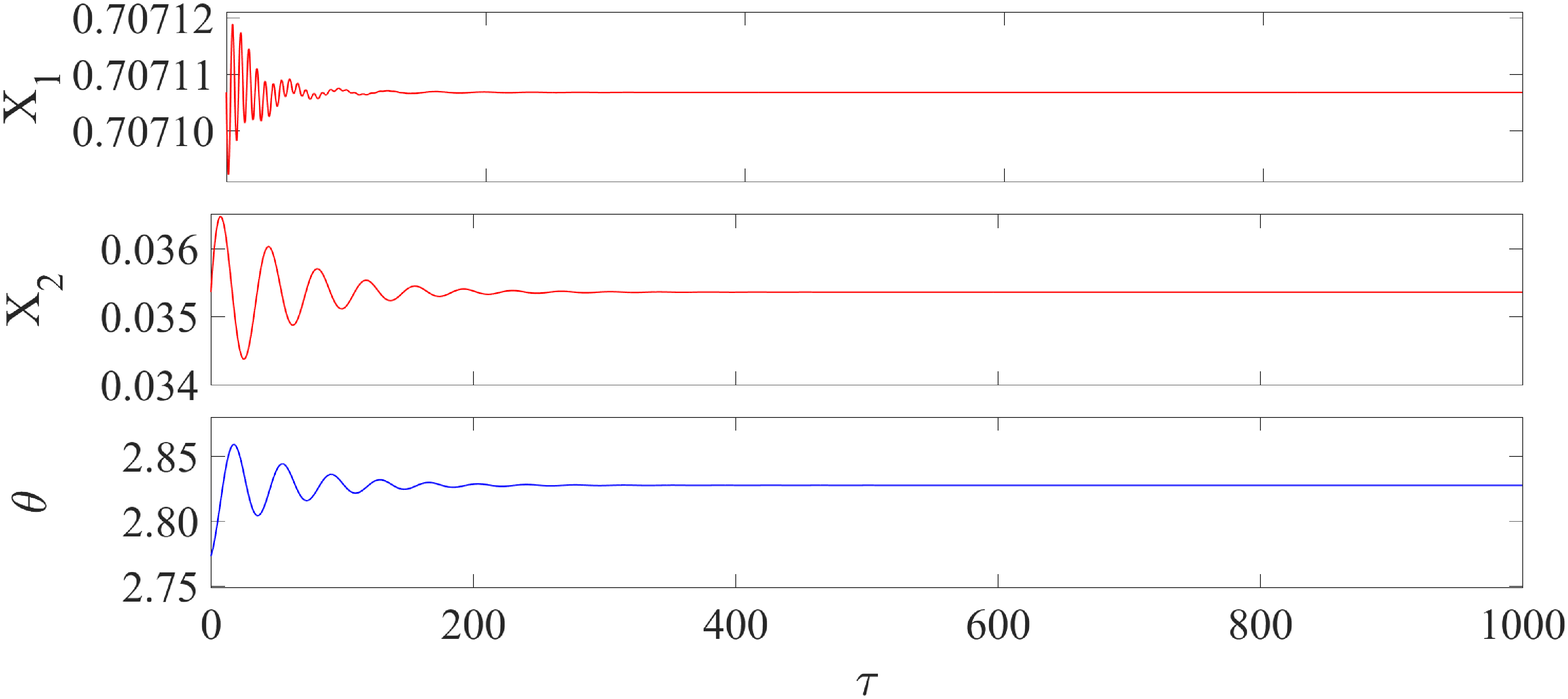}}}
   \caption{Time evolution of the electric field ampltudes and phase difference for parameters corresponding to Fig. 4(a). The initial conditions correspond to asymmetric phase-locked states with $\theta=0.9\pi\simeq2.83$ and $\rho=$0.75 (a), 0.50 (b), 0.25 (c), 0.15 (d), 0.10 (e), 0.05 (f), perturbed by random noise. In accordance to Fig. 4(a), the phase locked states are unstable for $0.06<\rho<0.54$. Cases of stable phase-locked states are shown in (a), (f). In the case of unstable phase-locked states the system evolves either to stale limit cycles [(b), (c), (e)] or to chaotic states (d). }
  \end{center}
\end{figure}

\section{Phased-locked states with arbitrary asymmetry under non-zero detuning}
Analogously to the case of zero detuning we can find analytically the steady-state carrier densities $(Z_{1,2})$ as well as the appropriate detuning $(\Delta)$ and pumping rates $(P_{1,2})$ for an arbitrary field amplitude ratio $(\rho)$ and phase difference $(\theta)$,  as follows    
\begin{eqnarray}
Z_1&=&\Lambda \rho \sin\theta \nonumber \\
Z_2&=&-\frac{\Lambda}{\rho}\sin\theta \label{Z_eq}
\end{eqnarray} 
\begin{eqnarray}
 \Delta&=&-\alpha \Lambda\sin\theta\left(\frac{1}{\rho}+\rho\right)-\Lambda\cos\theta\left(\frac{1}{\rho}-\rho\right)  \label{D_eq}\\
 P_1&=&X_0^2+(1+2X_0^2)\Omega\Lambda\rho\sin\theta  \nonumber\\
 P_2&=&\rho^2X_0^2-(1+2\rho^2 X_0^2) \frac{\Omega\Lambda}{\rho}\sin\theta  \label{P_eq}
\end{eqnarray}
Therefore, there always exists a phase-locked state with arbitrary field amplitude asymmetry and phase difference, provided that the detuning $\Delta$ and the pumping rates $P_{1,2}$ have values given by Eqs. (\ref{D_eq}) and (\ref{P_eq}), respectively, while the steady-state carrier densities $(Z_{1,2})$ are given by Eqs. (\ref{Z_eq}). These phase-locked states exist in the whole parameter space and can have an arbitrary power $X_0$. However, their stability depends strongly on the coupling $(\Lambda)$ as well as the power $(X_0)$ and the degree of their asymmetry, characterized by $\rho$ and $\theta$, as shown in Figs. 4(a)-(f). It is worth mentioning that there is enough freedom in parameter selection in order to have a controllable configuration that supports a large variety of stable asymmetric phase-locked states with unequal field amplitudes and phase differences, with the latter crucially determining the far field patterns of the pair of coupled lasers. The asymmetric states are characterized by carrier densities having opposite signs $Z_1/Z_2=-\rho^2<0$ so that the electric fields of the two lasers experience gain and loss, respectively. For $\rho=1$ we have equal gain and loss and a phase-locked state with equal field amplitude and phase difference given by Eq. (\ref{D_eq}) as $\sin\theta=-\Delta/2\alpha\Lambda$. At the boundaries of the stability regions, the system undergoes Hopf bifurcations giving rise to stable limit cycles characterized by asymmetric synchronized oscillations of the electric fields, that can have different mean values and amplitudes of oscillation. 

Characteristic cases for the time evolution of the electric field amplitudes $X_{1,2}$ and the phase difference $\theta$ are depicted in Fig. 5 for various degrees of asymmetry. The parameters of the system correspond to those of Fig. 4(a), with  phase difference $\theta=0.9\pi\simeq2.83$ and various values of $\rho$. For $\rho=0.75$ [Fig. 5(a)] the asymmetric phase-locked state is stable and perturbed initial conditions evolve to the stable state. As $\rho$ decreases to $\rho=0.5$ [Fig. 5(b)] and $0.25$ [Fig. 5(c)], the phase-locked states become unstable and the system evolves to stable limit cycles of increasing period. Close to the center of the unstable region the system evolves to chaotic states [$\rho=0.15$, Fig. 5(d)]. Further decreasing $\rho$ results in stable limit cycles [$\rho=0.10$, Fig. 5(e)] and stable phase-locked states [$\rho=0.05$, Fig. 5(f)] corresponding to the stability region of lower $\rho$ shown in Fig. 4(a).

\begin{figure}[pt]
  \begin{center}
  \subfigure[]{\scalebox{\scl}{\includegraphics{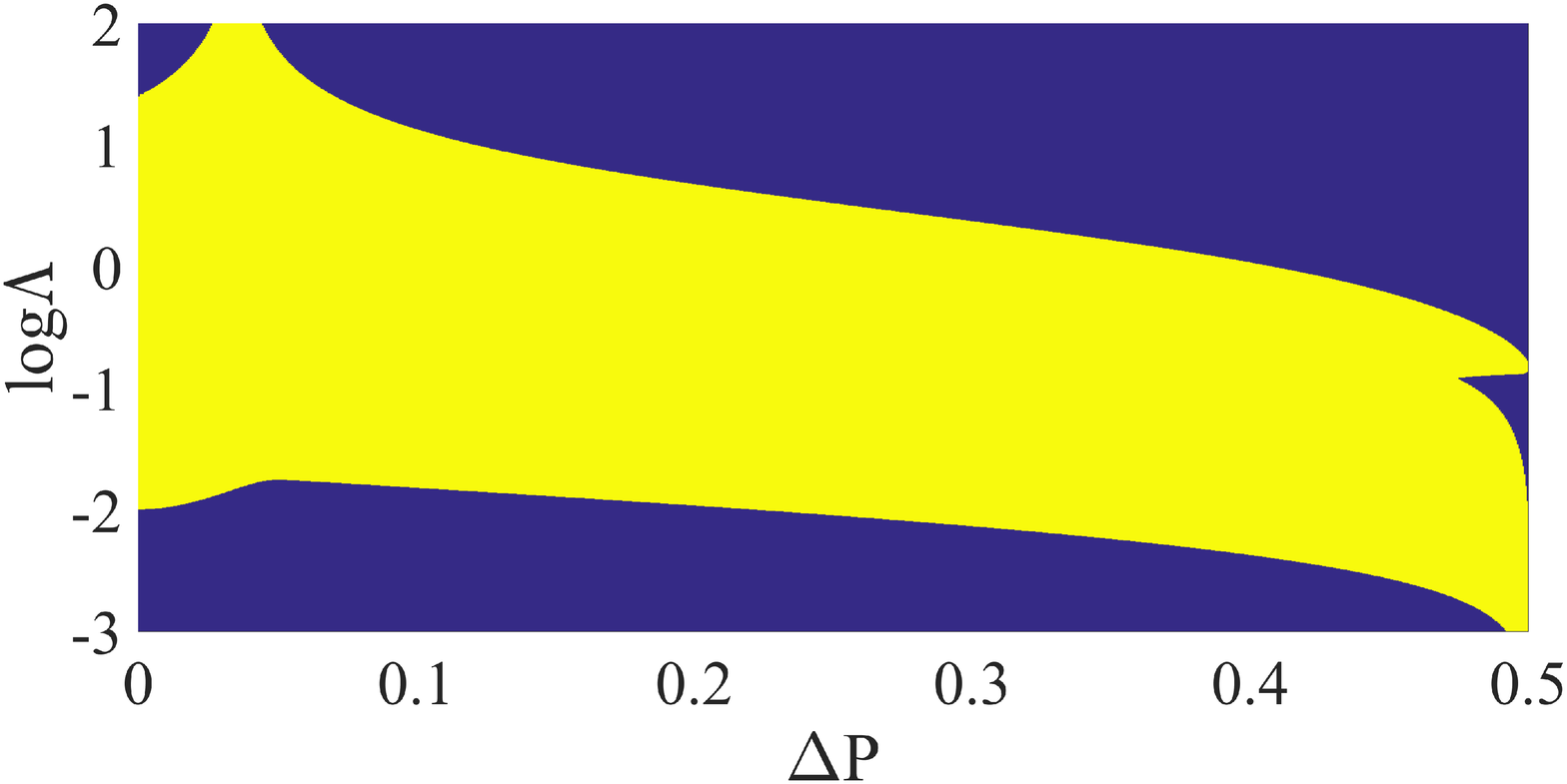}}}
  \subfigure[]{\scalebox{\scl}{\includegraphics{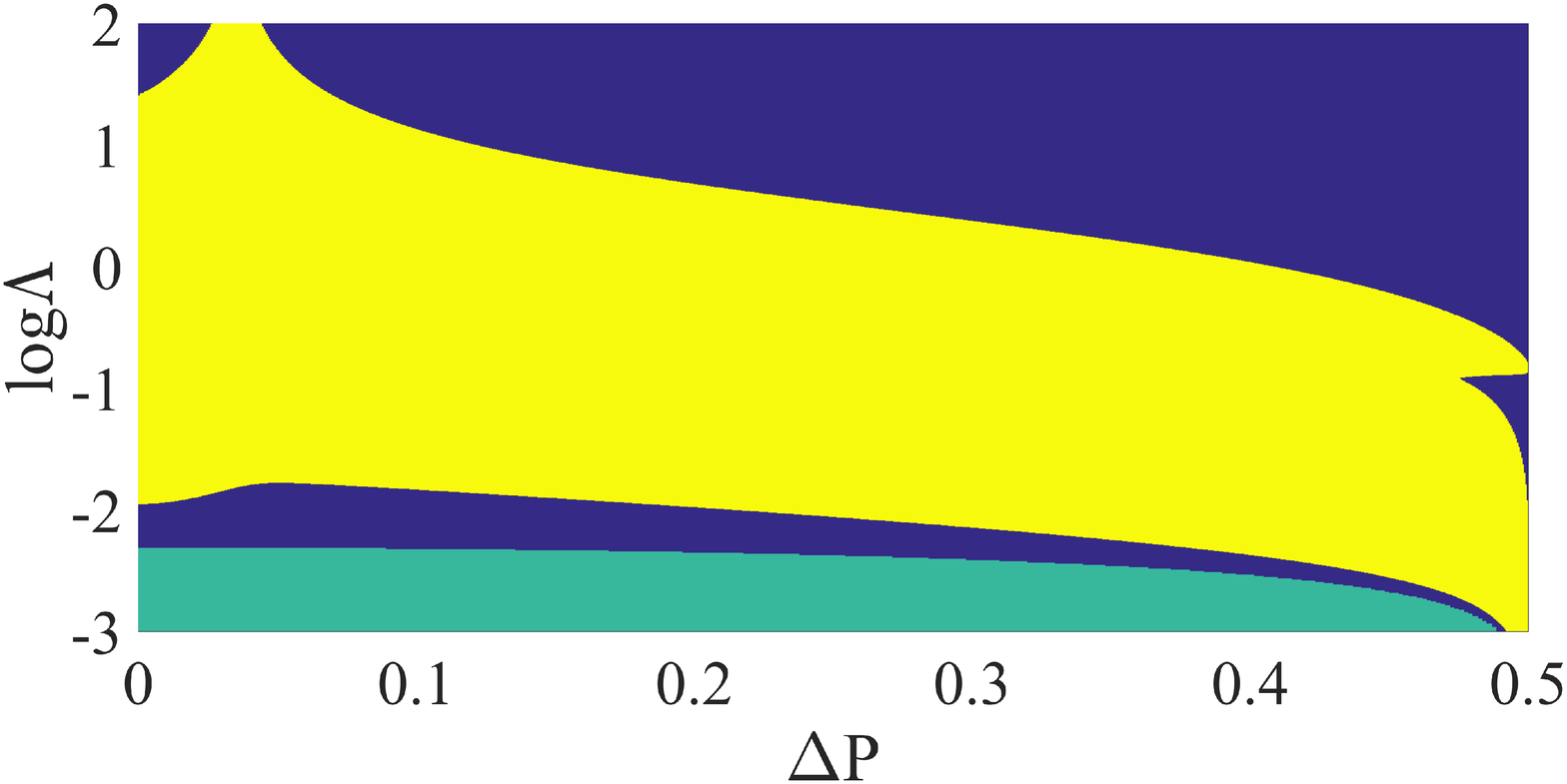}}}
  \subfigure[]{\scalebox{\scl}{\includegraphics{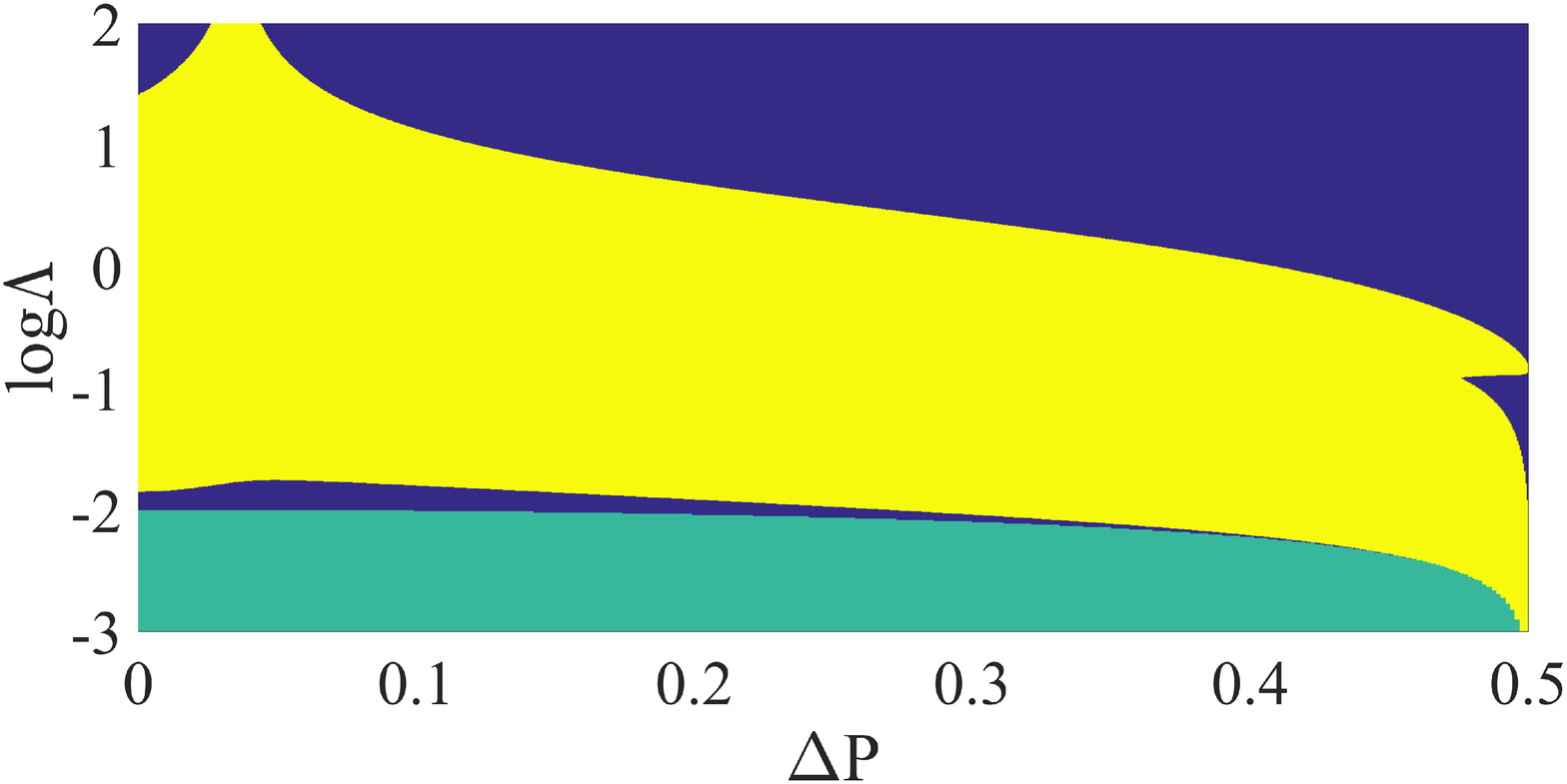}}}
  \caption{Existence and stability regions of phase-locked states in the $(\Lambda, \Delta P)$ parameter space for $\alpha=5$ and $T=400$ when $P_0=0.5$. Dark blue and light yellow areas correspond to stability and instability, respectively. The green area corresponds to nonexistence of a phased-locked state due to non-zero detuning. (a) $\Delta =0$, (b) $\Delta =0.05$, (c) $\Delta =0.1$. }
  \end{center}
\end{figure}

\section{Deformation of symmetric phased-locked states under nonzero detuning and pumping asymmetry}
 For the case of a given nonzero detuning and/or asymmetrically pumped lasers $P_1=P_0+\Delta P, P_2=P_0-\Delta P$, the equilibria of the system (\ref{pair}) cannot be analytically obtained for a given set of values $(\Delta, \Delta P)$. The respective algebraic system consists of transcendental equations and is solved by utilizing a numerical continuation algorithm, according to which we start from $\Delta=0$ and $\Delta P=0$ corresponding to the symmetric case with the two known equilibria $(\theta=0,\pi)$. For each one of them, we increase $\Delta$ and/or $\Delta P$ in small steps; in each step the solution of the previous step is used as an initial guess for the iterative procedure (Newton-Raphson method) that provides the solution. 

For the case of zero detuning $\Delta=0$, the domain of existence of stable phase-locked states in the ($\Lambda$, $\Delta P$) parameter space is shown in Fig. 6(a). For $\Delta P =0$ the results are similar to the case considered in \cite{Winful&Wang_88}, with the in-phase state being stable for large $\Lambda$ and the out-of-phase being stable for small values of $\Lambda$. As the pumping difference increases, the stable in-phase state extends only over a quite small range of $\Delta P$, whereas the out-of-phase state extends almost in the entire range of $\Delta P$. In both cases, as $\Delta P$ increases from zero the phase difference is slightly differentiated from the values $\theta=0,\pi$ for $\Delta P = 0$. Surprisingly, another region of stable phase-locked states appears in the strong coupling regime (large $\Lambda$) above a threshold of pumping difference $\Delta P$. This is an out-of-phase state with phase difference close to $\pi$ that appears for values of $\Delta P$ for which no stable in-phase state exists. In fact, this stable state exists for a large part of the parameter space and extends to values $\Delta P=P_0$ for which only one of the lasers is pumped above threshold ($P_1=1$, $P_2=0$). This area of stability extending from intermediate to high values of coupling is enabled by the asymmetric pumping and indicates its stabilizing effect.   
The dependence of the phase difference of the stable states on the coupling $\Lambda$ and pumping difference $\Delta P$ is depicted in Fig. 7(a). In Fig. 8, the electric field amplitudes $X_{1,2}$ and carrier densities $X_{1,2}$ of all the stable states are shown. It is clear that for $\Delta P=0$ we have $X_{1,2}=\sqrt{P_0}$ and $Z_{1,2}=0$ and for small values of $\Lambda$ we have $X_{1,2}=\sqrt{P_{1,2}}$ and $Z_{1,2}=0$, as the two lasers are essentially uncoupled. For  $\Delta P>0$ and finite coupling values the electric field and carrier density destributions in the two lasers become highly asymmetric.        
 \begin{figure}[pt]
  \begin{center}
  \subfigure[]{\scalebox{\scl}{\includegraphics{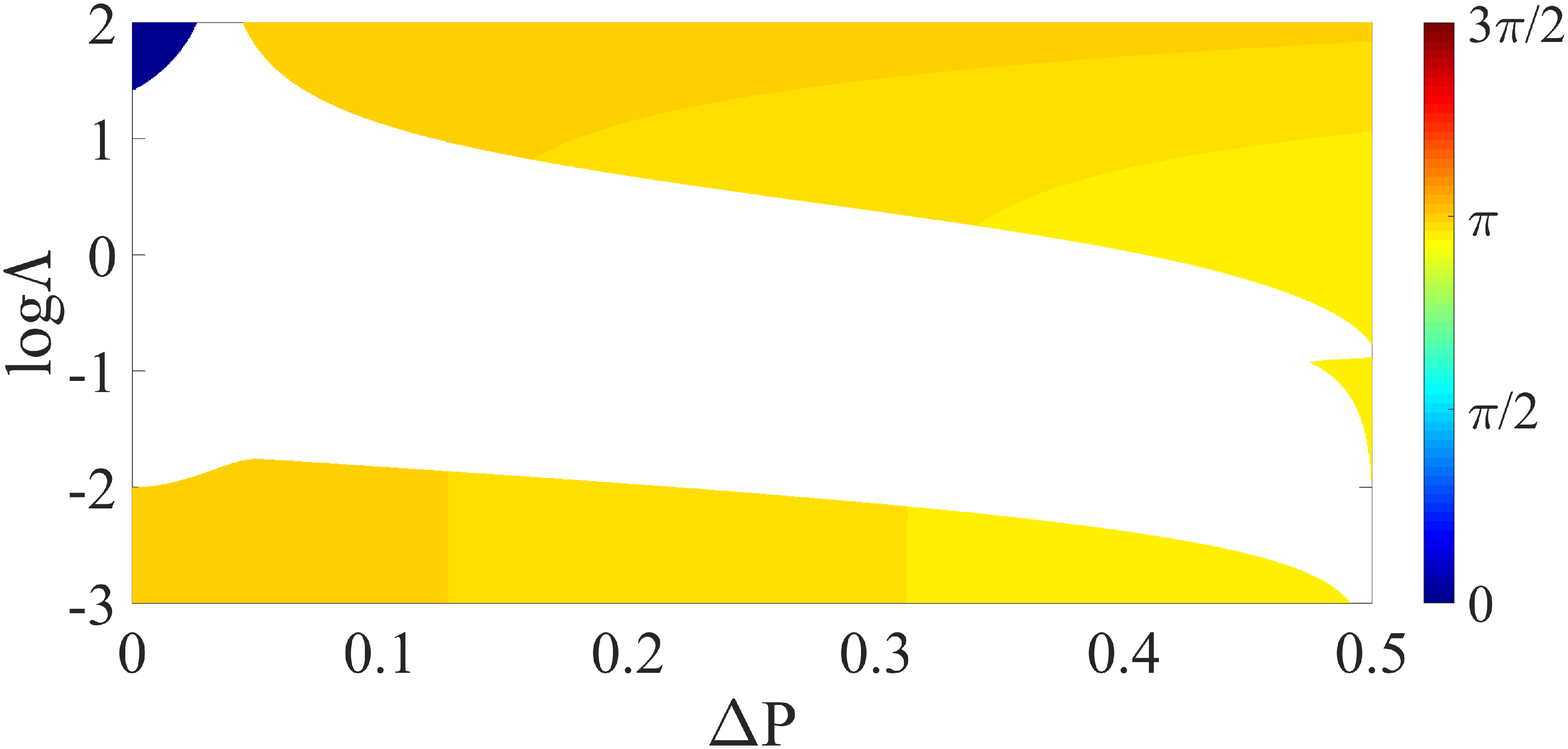}}}
  \subfigure[]{\scalebox{\scl}{\includegraphics{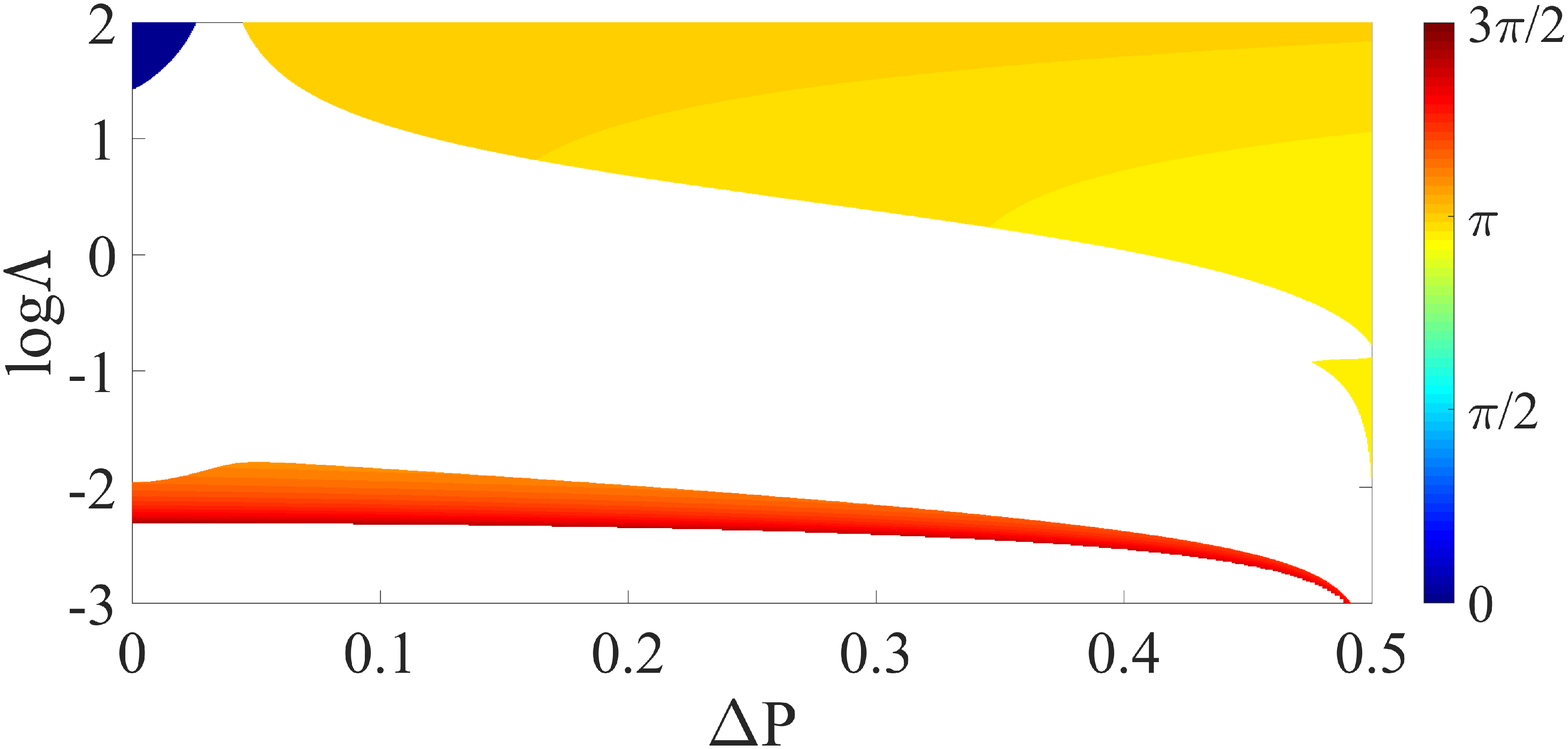}}}
  \caption{Steady-state phase difference ($\theta$) of the stable phase-locked states in the $(\Lambda, \Delta P)$ parameter space for $\alpha=5$, $T=400$ and $P_0=0.5$. (a) $\Delta=0$, (b) $\Delta=0.05$ correspond to the cases of Fig. 6(a) and (b), respectively. }
  \end{center}
\end{figure}

\begin{figure}[pt]
  \begin{center}
  \subfigure[]{\scalebox{\scl}{\includegraphics{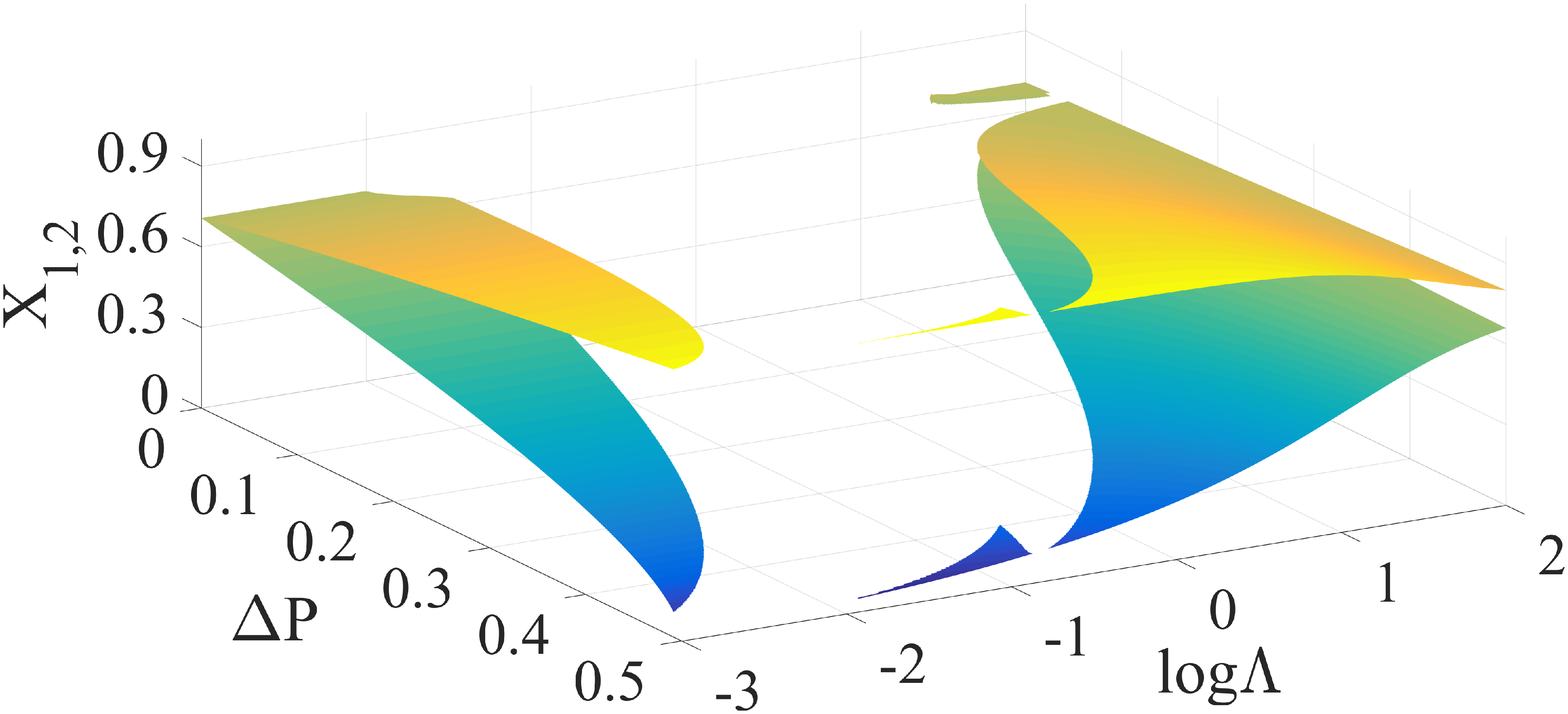}}}
  \subfigure[]{\scalebox{\scl}{\includegraphics{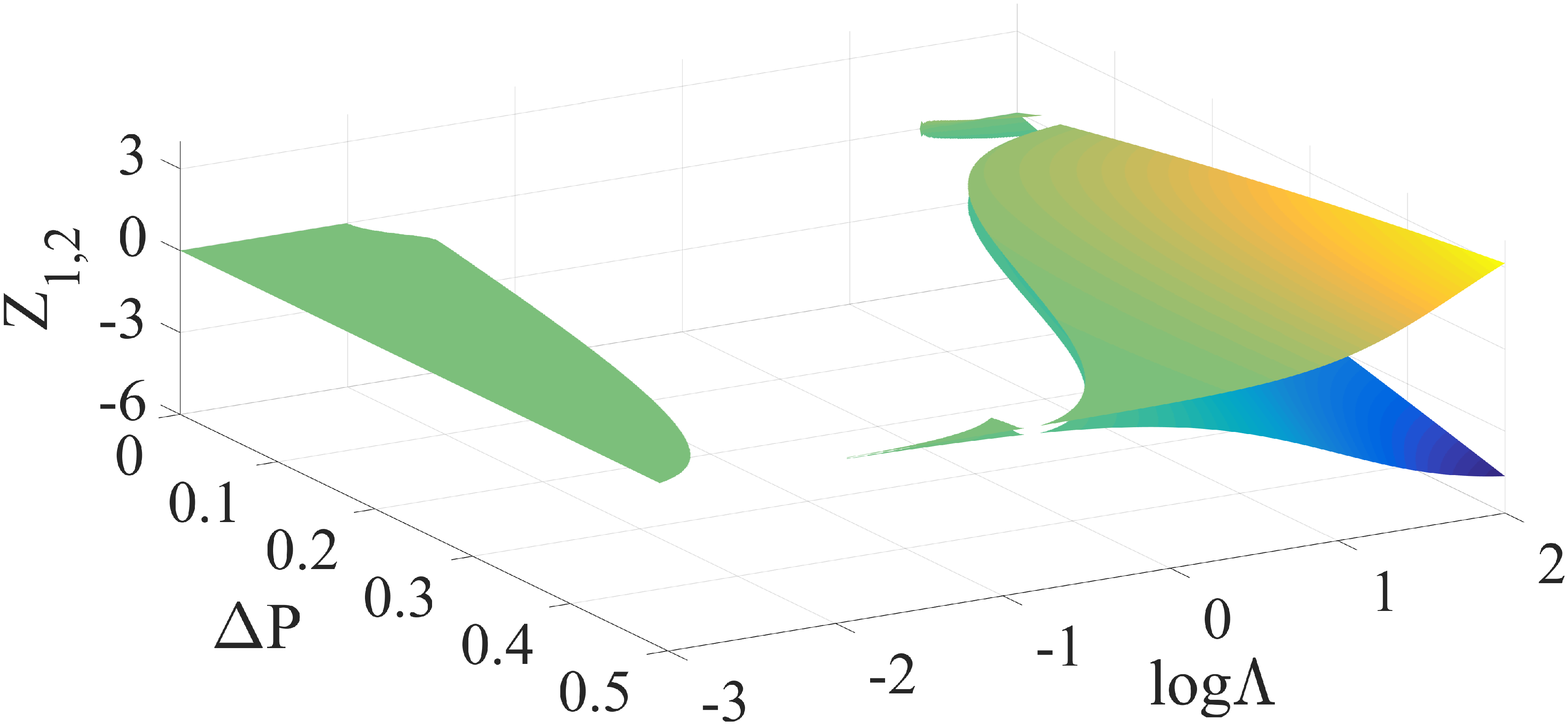}}}
  \caption{Steady-state electric field amplitudes $X_{1,2}$ (a) and carrier densities $Z_{1,2}$ (b) of the stable phase-locked states in the $(\Lambda, \Delta P)$ parameter space for $\alpha=5$, $T=400$ and $P_0=0.5$, corresponding to the case of Fig. 6(a). }
  \end{center}
\end{figure}

\begin{figure}[pt]
  \begin{center}
  \subfigure[]{\scalebox{\scl}{\includegraphics{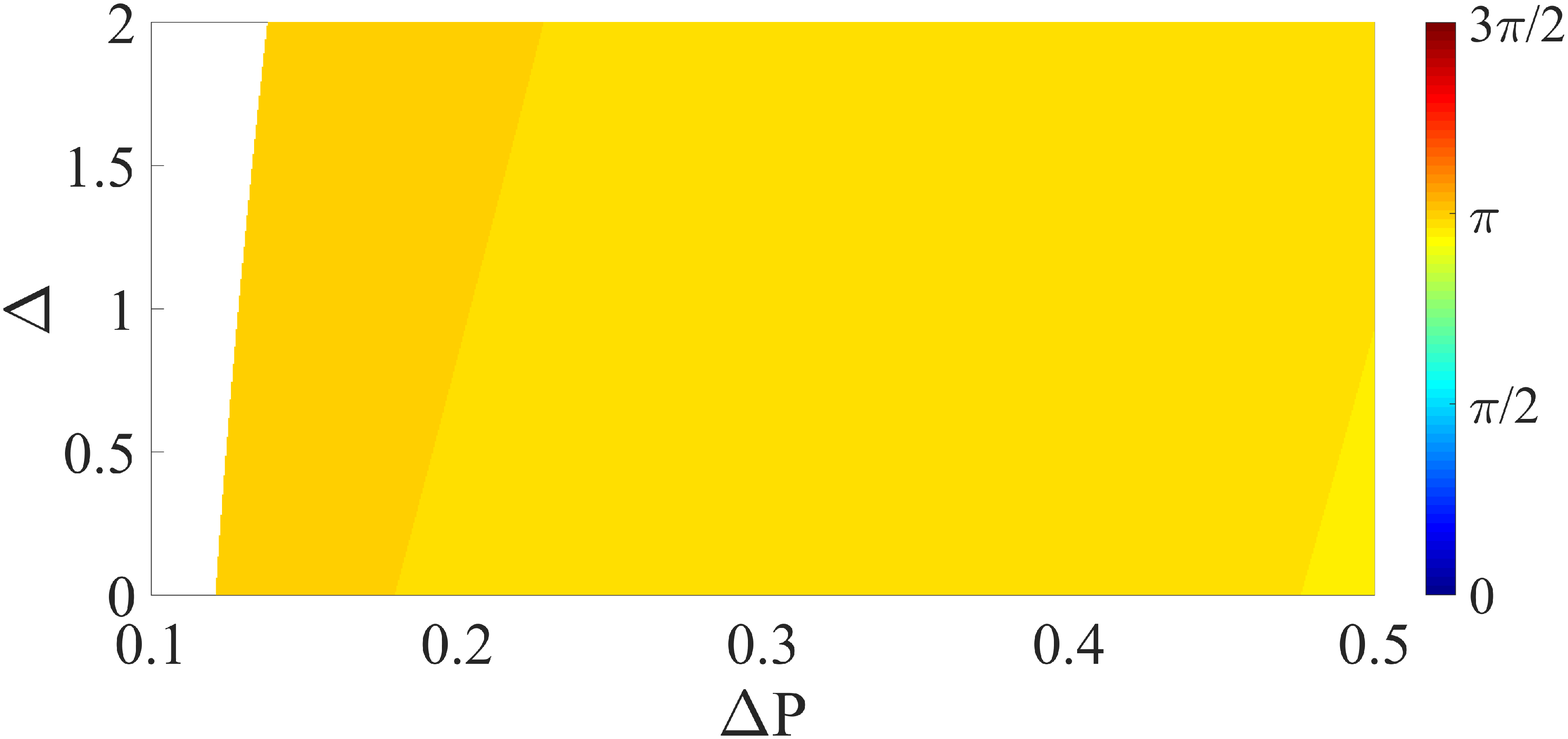}}}
  \subfigure[]{\scalebox{\scl}{\includegraphics{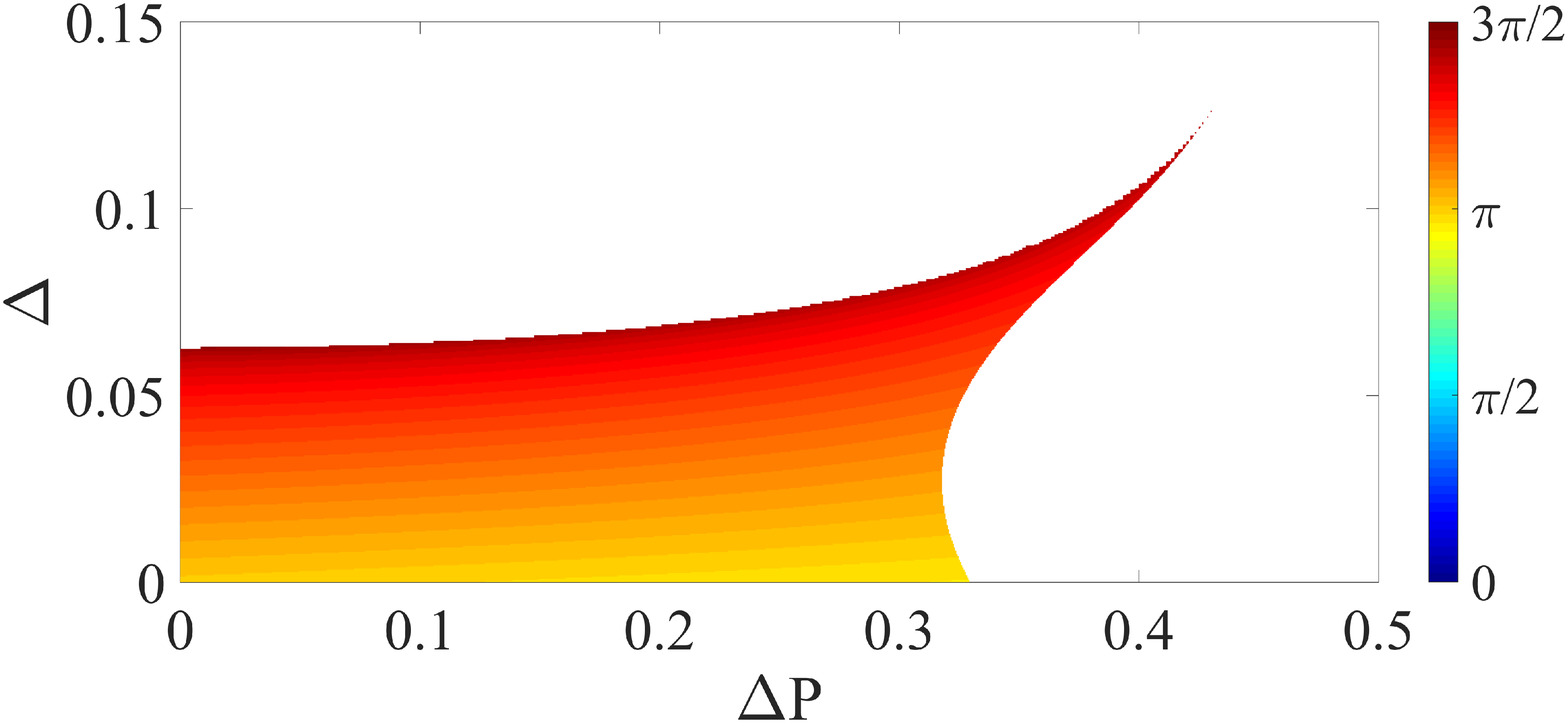}}}
   \caption{Steady-state phase difference ($\theta$) of the stable phase-locked states in the $(\Delta, \Delta P)$ parameter space for $\alpha=5$, $T=400$ and $P_0=0.5$. (a) strong coupling $\log \Lambda=1$, (b) weak coupling $\log \Lambda =-2.2$.  }
  \end{center}
\end{figure}

A non-zero detuning ($\Delta \neq 0$) between the two lasers strongly affects the existence of a stable out-of-phase state in the weak coupling regime, as shown in Figs. 6(b), (c)  for $\Delta=0.05, 0.1$, respectively. The role of detuning in terms of the phase of the stable states is clearly presented in Fig. 7(b) for $\Delta=0.05$. In comparison to Fig. 7(a), corresponding to zero detuning, it is obvious that the phase of the stable states existing for intermediate and strong coupling is hardly affected by the detuning whereas the stable state, existing in the weak coupling regime, has a phase that ranges from $\pi$ to $3\pi/2$ depending strongly on $\Lambda$ and $\Delta P$. The role of detuning is also shown in Fig. 9(a) and (b) for $\Delta =0.05$ in the case of strong ($\log \Lambda=1$) and weak ($\log \Lambda =-2.2$) coupling, respectively. The detuning introduces phase sensitivity to current injection for stable modes of the weak coupling regime that can be quite interesting for beam-steering applications \cite{Choquette_13}.

\section{Conclusions}
We have investigated the existence of stable asymmetric phase-locked states in a system of two coupled semiconductor lasers. The asymmetric phase-locked states are characterized by a non-unitary field amplituded ratio and non-trivial phase difference. The asymmetry is shown to be directly related to operation conditions that result in the presence of gain in one laser and loss in the other. The crucial role of carrier density dynamics has been taken into account by considering a model where the field equations are dynamically coupled with the carrier density equations, in contrast to standard coupled mode equations, commonly considered in studies on non-Hermitian photonics and PT-symmetric lasers. It has been shown that stable asymmetric states exist even in absolutely symmetric configurations and that states of arbitrary asymmetry can be supported by appropriate selection of the detuning and the pumping profile of the system. The role of the current injection suggests a dynamic mechanism for the control of the phase-locked states and, therefore, the far-field emission patterns of this fundamental photonic element consisting of two coupled lasers.

\section*{Acknowledgements}
Y. K. is grateful to the School of Science and Technology of Nazarbayev University, Astana, Kazakhstan for its hospitality during his visit at NU in November, 2016. This research is partly supported by the ORAU grant entitled "Taming Chimeras to Achieve the Superradiant Emitter", funded by Nazarbayev University, Republic of Kazakhstan. This work was partially also supported by the Ministry of Education and Science of the Republic of Kazakhstan via Contract No. 339/76-2015.

\end{document}